\documentclass[preprint]{aastex}
\slugcomment{accepted for publication in Astrophysical Journal}
\shorttitle{Emission line ratios in five edge--on galaxies}
\shortauthors{B. Otte et al.}

\begin{document}

\title{Emission Line Ratios and Variations in Temperature and Ionization State
in the Diffuse Ionized Gas of Five Edge--on Galaxies}
\author{B. Otte\altaffilmark{1,2}, J. S. Gallagher III\footnotemark[1], R. J.
Reynolds}
\footnotetext[1]{Visiting Astronomer, Kitt Peak National Observatory, National
Optical Astronomy Observatories, which is operated by the Association of
Universities for Research in Astronomy, Inc. (AURA) under cooperative agreement
with the National Science Foundation.}
\altaffiltext{2}{Department of Physics and Astronomy, Johns Hopkins University,
3400 North Charles Street, Baltimore, MD 21218}
\affil{Department of Astronomy, University of Wisconsin--Madison}
\affil{475 North Charter Street, Madison, WI 53706}
\email{otte@pha.jhu.edu, jsg@astro.wisc.edu, reynolds@astro.wisc.edu}

\begin{abstract}
We present spectroscopic observations of ionized gas in the disk--halo regions
of five edge--on galaxies, covering a wavelength range from
[\ion{O}{2}]\,$\lambda$3727\AA\ to [\ion{S}{2}]\,$\lambda$6716.4\AA. The
inclusion of the [\ion{O}{2}] emission provides additional constraints on the
properties of the diffuse ionized gas (DIG), in particular, the origin of the
observed spatial variations in the line intensity ratios. We have derived
electron temperatures, ionization fractions and abundances along the slit. Our
data include both slit positions parallel and perpendicular to the galactic
disks. This allowed us to examine variations in the line intensity ratios with
height above the midplane as well as distance from the galactic centers. The
observed increase in the [\ion{O}{2}]/H$\alpha$ line ratio towards the halo
seems to require an increase in electron temperature caused by a non--ionizing
heating mechanism. We conclude that gradients in the electron temperature can
play a significant role in the observed variations in the optical emission line
ratios from extraplanar DIG.
\end{abstract}

\keywords{ISM: abundances --- ISM: general --- ISM: individual (NGC\,891,
NGC\,3079, NGC\,4631, NGC\,4634, NGC\,5775) --- galaxies: abundances ---
galaxies: general --- galaxies: individual (NGC\,891, NGC\,3079, NGC\,4631,
NGC\,4634, NGC\,5775)}

\section{INTRODUCTION}

Extraplanar diffuse ionized gas (eDIG) has been observed in many galaxies in
addition to the \objectname[]{Milky Way} over the past ten years. However, the
origin of this gas, and how it is heated and ionized up to several kiloparsecs
away from hot OB stars in the midplane of the galaxies is still not well
understood. Many attempts have been made to explain the eDIG by pure
photoionization models (e.g. Domg\"orgen \& Mathis 1994) or combinations of
photoionization and at least one additional ionization process (e.g. Martin
1997; Collins \& Rand 2001). This additional process was needed to explain the
rise in the [\ion{O}{3}]/H$\beta$ line ratio with increasing distance from the
disk. However, even with these composite models it was not possible to explain
the constant [\ion{S}{2}]/[\ion{N}{2}] line ratio, which was observed in
\objectname[]{NGC\,891} \citep{2rand}, as well as in the \objectname[]{Milky
Way} (Haffner, Reynolds, \& Tufte 1999) and other galaxies \citep{2otte99}.
These data led \citet{2haffner} to the conclusion that the electron temperature
increases with increasing distance from the midplane of the galaxies. Such a
variation in temperature would seem to require the existence of an additional,
non--ionizing heating process within the DIG (Reynolds, Haffner, \& Tufte 1999).

A rise in temperature can explain both the growing [\ion{O}{3}]/H$\beta$ ratio
as well as the constant [\ion{S}{2}]/[\ion{N}{2}] ratio with increasing galactic
altitude $|z|$ without invoking an additional ionization mechanism at high
$|z|$. Such a rise in electron temperature also would affect the
[\ion{O}{2}]/H$\alpha$ line ratio, and because of its relatively high excitation
energy, the [\ion{O}{2}]\,$\lambda$3727\AA\ emission line provides important
additional information about the ionization and heating processes in the DIG. In
an earlier study, we measured line intensity ratios, including the [\ion{O}{2}]
emission line, as a function of height above the midplane in three edge--on
galaxies, using three different analysis methods (Otte et al. 2001, hereafter
Paper I). Paper I shows that intensity ratios of the brighter optical emission
lines from the eDIG can be understood, if the electron temperature tends to
increase with height above the galactic midplane. In this paper, we examine the
same galaxies, but at slit positions parallel to the midplane, and two
additional galaxies, using the method that yielded the most reasonable results
in Paper I. The new results confirm our previous conclusion, in which case
photoionization would not be the sole energy source for the eDIG.

\section{OBSERVATIONS AND DATA REDUCTION}

\begin{figure}
\plotone{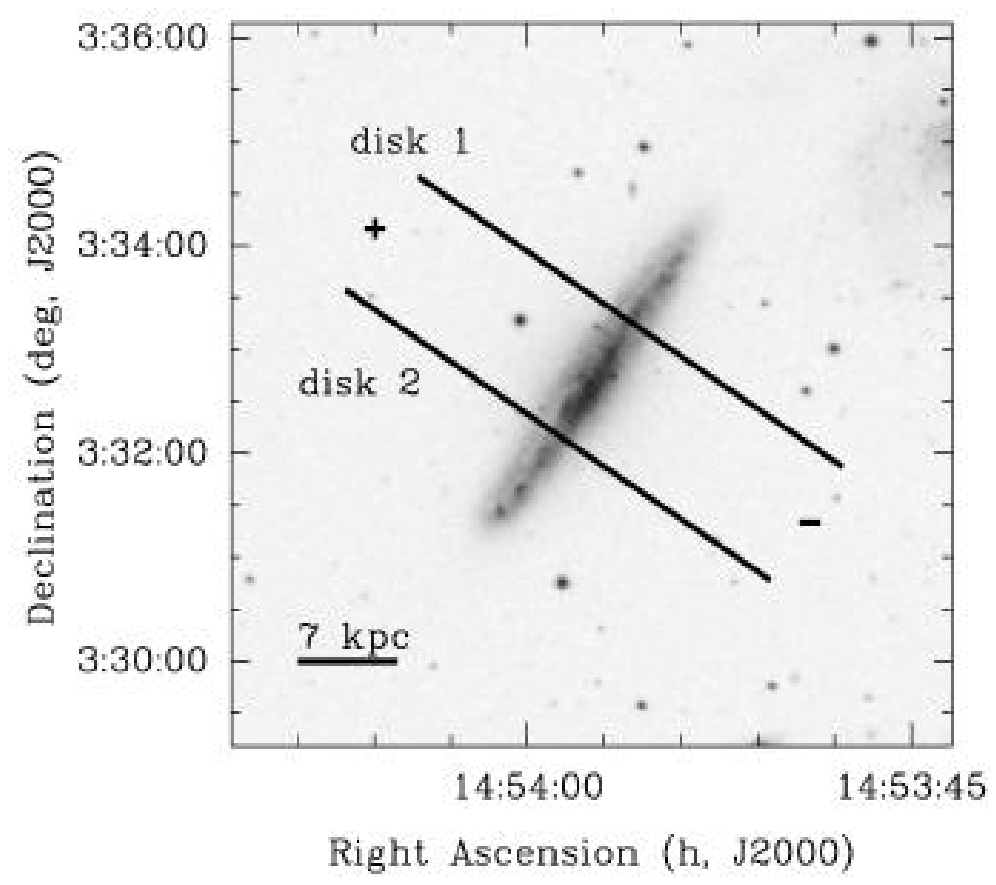}
\caption{\label{2f1}
NGC\,5775 taken from the Digitized Sky Survey (Second Generation). The positions
of the disk slits are shown. The $+$ and $-$ signs indicate the counting
direction of the height $z$ in our plots. We assumed a distance of 24.8\,Mpc.}
\end{figure}

\begin{figure}
\epsscale{0.8}
\plotone{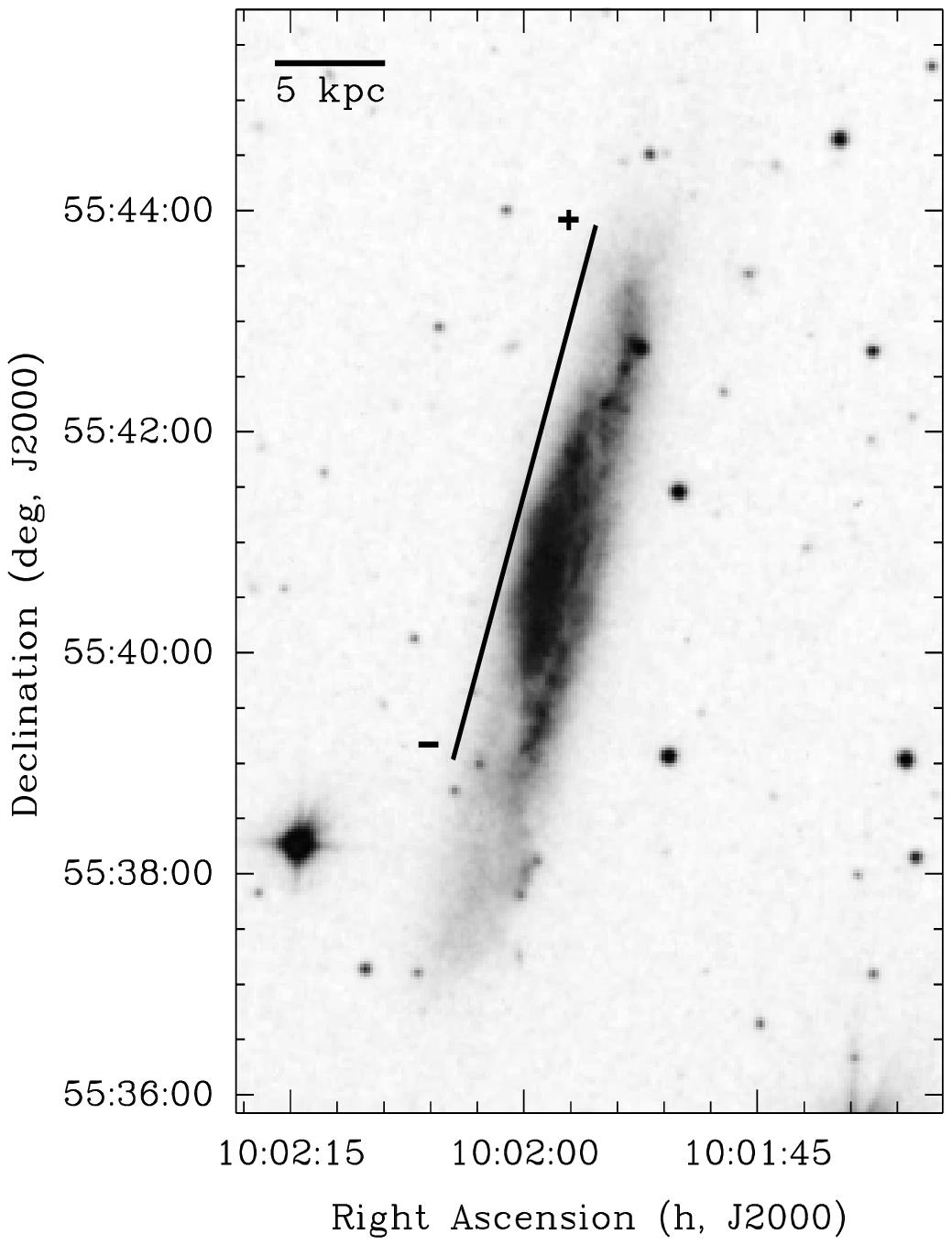}
\caption{\label{2f2}
NGC\,3079 taken from the Digitized Sky Survey (First Generation). The position
of the DIG slit is shown. The $+$ and $-$ signs indicate the counting direction
of the radius $r$ in our plots. We assumed a distance of 17.3\,Mpc.}
\end{figure}

\begin{figure}
\epsscale{1.0}
\plotone{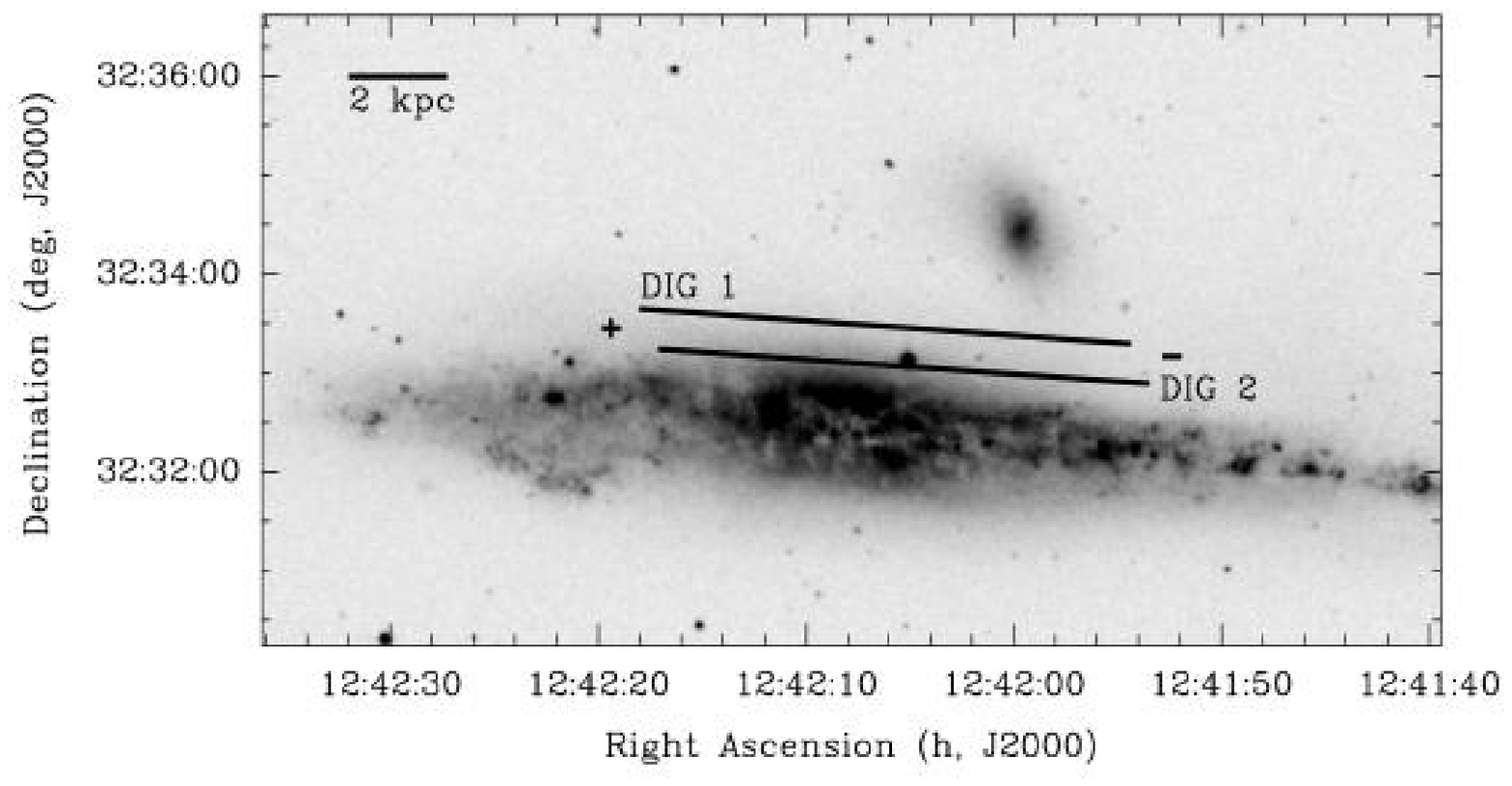}
\caption{\label{2f3}
NGC\,4631 taken from the Digitized Sky Survey (Second Generation). The positions
of the DIG slits are shown. The $+$ and $-$ signs indicate the counting
direction of the radius $r$ in our plots. We assumed a distance of 6.9\,Mpc.}
\end{figure}

\begin{figure}
\plotone{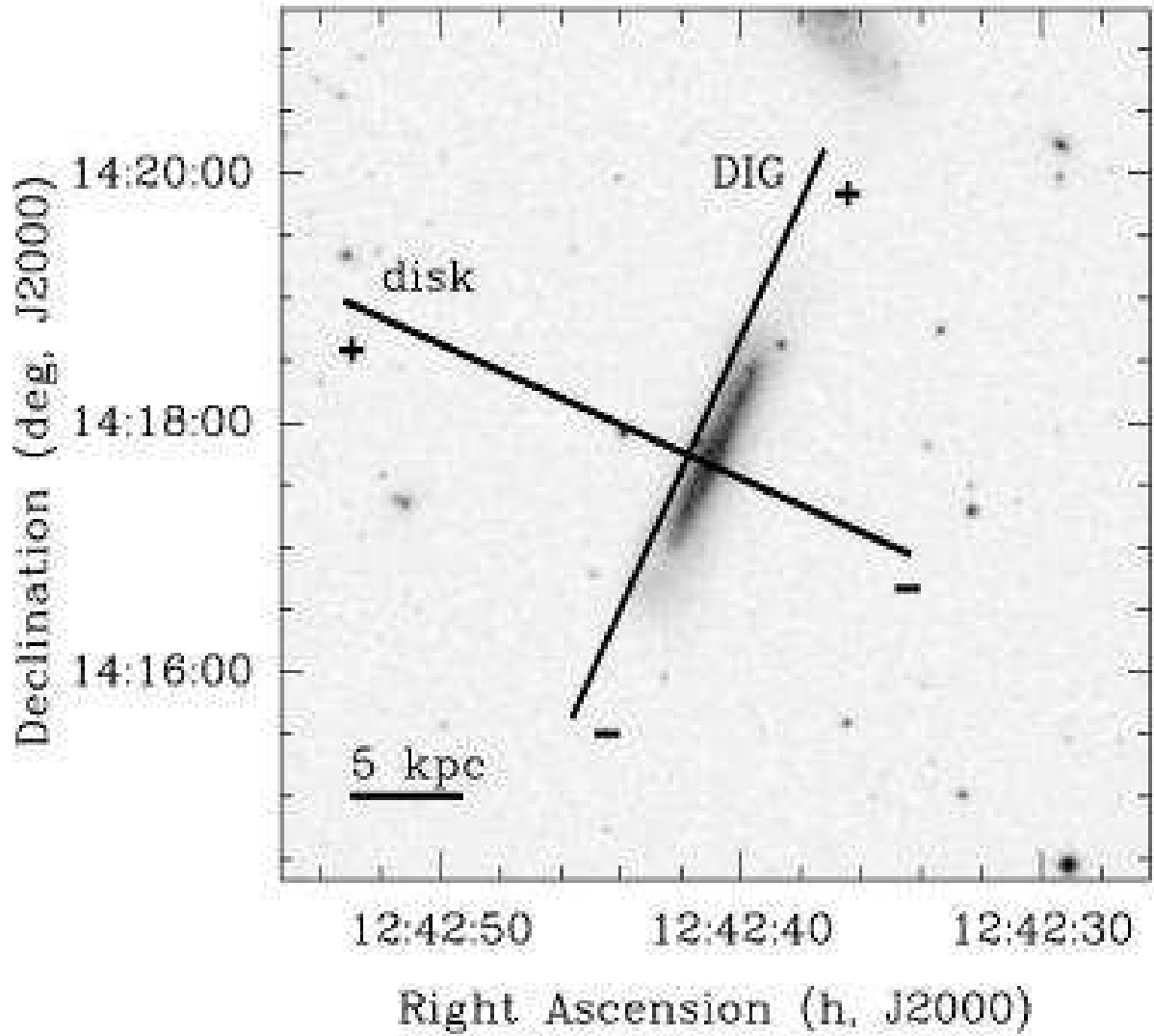}
\caption{\label{2f4}
NGC\,4634 taken from the Digitized Sky Survey (Second Generation). The positions
of the slits are shown. The $+$ and $-$ signs indicate the counting direction of
the radius $r$ or the height $z$ in our plots. We assumed a distance of
19\,Mpc.}
\end{figure}

\begin{figure}
\plotone{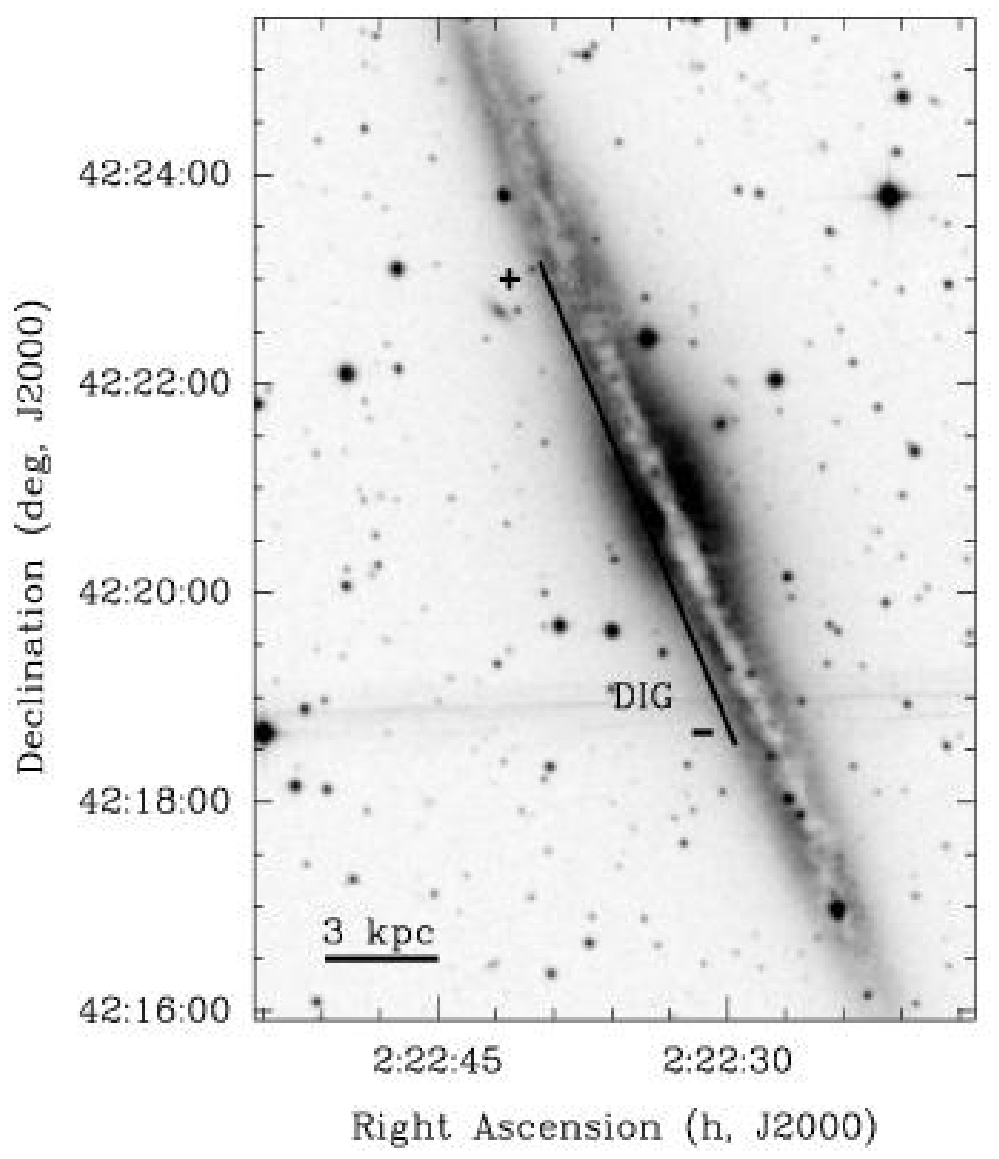}
\caption{\label{2f5}
NGC\,891 taken from the Digitized Sky Survey (Second Generation). The position
of the DIG slit is shown. The $+$ and $-$ signs indicate the counting direction
of the radius $r$ in our plots. We assumed a distance of 9.6\,Mpc.}
\end{figure}

All spectra were obtained with the GoldCam spectrograph at the 2.1\,m telescope
on Kitt Peak, AZ, in 2000 February 29 -- March 6 and September 26 -- 27. We used
grating 9 with decker 4 and a slitwidth of 2$\farcs$5. This yielded a pixel
scale of 0$\farcs$80/pixel or 2.44\,\AA/pixel, respectively, and a wavelength
range from about 3500\,\AA\ to 7400\,\AA. The spectral resolution was dominated
by the instrumental line width and ranges from 410 at [\ion{O}{2}] to 750 at
[\ion{S}{2}]. We used filter WG345 to remove possible overlaps between orders.
Exposure times varied from 20\,min to 30\,min depending on the weather (with the
shorter exposure times during the first night which was partly cloudy). Table
\ref{2t1} lists the number of combined exposures, the total integration time for
each slit position, the distance of the slit position from the galactic center,
and the scale. The slit positions are shown in Figures \ref{2f1} to \ref{2f5}.
``DIG'' slits are parallel to the disk, and ``disk'' slits are perpendicular
through the disk of the galaxies. We also took spectra of blank sky regions
before or after each set of galaxy observations. The emission lines at the
parallel slit position at higher galactic latitude in \objectname[]{NGC\,4631}
(DIG 1 slit) were too weak, so that we moved the slit closer to the midplane of
the galaxy (DIG 2 slit). We could not identify any [\ion{O}{2}] emission in the
raw data of \objectname[]{NGC\,5775} and therefore abandoned this galaxy during
our observing run. However, the [\ion{O}{2}] emission line became visible after
careful sky subtraction.
\begin{deluxetable}{llcccc}
\tablewidth{0pt}
\tablecaption{\label{2t1} OBSERVATIONS}
\tablehead{\colhead{Object} & \colhead{Slit Position} & \colhead{Exposures} &
\colhead{Total Integration Time} & \colhead{Distance to Center\tablenotemark{a}}
& \colhead{Scale} \\
\colhead{} & \colhead{} & \colhead{} & \colhead{(min)} & \colhead{(kpc)} &
\colhead{(arcsec kpc$^{-1}$)}}
\startdata
\objectname[]{NGC\,891} & DIG & 3 & 90 & 0.7 & 21.5 \\
\objectname[]{NGC\,3079} & DIG & 9 & 270 & 2.3 & 11.9 \\
\objectname[]{NGC\,4631} & DIG 1 & 2 & 60 & 2.0 & 29.9 \\
\objectname[]{NGC\,4631} & DIG 2 & 9 & 265 & 1.2& 29.9  \\
\objectname[]{NGC\,4634} & disk & 4 & 110 & 0.5 & 10.9 \\
\objectname[]{NGC\,4634} & DIG & 3 & 90 & 1.0 & 10.9 \\
\objectname[]{NGC\,5775} & disk 1 & 1 & 20 & 4.8 & 8.3 \\
\objectname[]{NGC\,5775} & disk 2 & 2 & 40 & 4.4 & 8.3 \\
\enddata
\tablenotetext{a}{Center defined by the NASA/IPAC Extragalactic Database (NED)}
\end{deluxetable}

The spectra were reduced using standard procedures in IRAF\footnote{IRAF is
distributed by the National Optical Astronomy Observatories, which are operated
by the Association of Universities for Research in Astronomy, Inc., under
cooperative agreement with the National Science Foundation.}. The analysis was
done in MIDAS\footnote{MIDAS is developed and maintained by the European
Southern Observatory.}. The line ratios were calculated from Gaussian fits to
the emission lines. Before measuring the emission lines, we averaged over
several rows to increase the signal--to--noise ratio. The number of averaged
rows and the corresponding interval for each slit position are listed in Table
\ref{2t2}. We point out that we used a gliding average. Therefore adjacent
data points are strongly correlated in the subsequent figures. However, we
believe that oversampling is necessary to show the full extent of the data. The
plotted error bars represent the statistical uncertainties of the smoothed
spectra. The H$\beta$ measurements were corrected for the underlying Balmer
absorption line. The high values of the H$\alpha$/H$\beta$ line ratio in the
plane of \objectname[]{NGC\,4634} and \objectname[]{NGC\,5775} (disk slits) show
that extinction correction is necessary for these slit positions. However, the
extinction correction seems to overcorrect the measurements in the blue part of
the spectrum. We therefore show both extinction corrected line ratios and
uncorrected line ratios for the disk slits in our analysis in section \ref{2s3}.
None of the DIG slit positions in our small sample shows a significant increase
in H$\alpha$/H$\beta$ over the theoretically expected value of 2.9 (for
10\,000\,K gas). We therefore did not apply an extinction correction to these
slit positions. A detailed description of the data reduction procedures and the
corrections for Balmer absorption lines and extinction can be found in Paper I.
\begin{deluxetable}{llcc}
\tablewidth{0pt}
\tablecaption{\label{2t2} EMISSION LINE AVERAGES}
\tablehead{\colhead{Object} & \colhead{Slit Position} & \colhead{Average} &
\colhead{Blur Factor\tablenotemark{a}} \\
\colhead{} & \colhead{} & \colhead{(rows)} & \colhead{(pc)}}
\startdata
\objectname[]{NGC\,891} & DIG & 9 & 300 \\
\objectname[]{NGC\,3079} & DIG & 5 & 270 \\
\objectname[]{NGC\,4631} & DIG 1 & 15 & 375 \\
\objectname[]{NGC\,4631} & DIG 2 & 9 & 215 \\
\objectname[]{NGC\,4634} & disk, $|z|\leq700$\,pc & 1 & 0 \\
\objectname[]{NGC\,4634} & disk, $|z|>700$\,pc & 9 & 590 \\
\objectname[]{NGC\,4634} & DIG & 9 & 590 \\
\objectname[]{NGC\,5775} & disk 1 & 9 & 770 \\
\objectname[]{NGC\,5775} & disk 2 & 9 & 770 \\
\enddata
\tablenotetext{a}{Number of averaged rows minus 1 converted to parsecs to show
the range in the subsequent figures that is affected by the averaging}
\end{deluxetable}

\section{\label{2s3} RESULTS}

Only four of the eight slit positions yielded reliable measurements for the six
emission lines necessary for our analysis ([\ion{O}{2}]\,$\lambda$3727\AA,
H$\beta$, [\ion{O}{3}]\,$\lambda$5006.9\AA, H$\alpha$,
[\ion{N}{2}]\,$\lambda$6583.4\AA, [\ion{S}{2}]\,$\lambda$6716.4\AA). Due to the
low resolution in our spectra, we were not able to distinguish between
[\ion{O}{2}]\,$\lambda$3726.0\AA\ and [\ion{O}{2}]\,$\lambda$3728.8\AA. Thus, we
use [\ion{O}{2}]\,$\lambda$3727\AA\ to mean the sum of both [\ion{O}{2}] doublet
lines. Our goal was to fit the line ratios [\ion{O}{2}]/H$\alpha$,
[\ion{O}{3}]/H$\alpha$, [\ion{N}{2}]/H$\alpha$ and [\ion{S}{2}]/H$\alpha$
keeping the number of assumptions to a minimum. Since each line ratio depends on
element abundance, ionization fraction and temperature, we had to make five
assumptions: 1) We assumed that all the emitting gas is ionized. This means in
particular that H$^+/{\rm H}=1$. 2) Since the ionization potentials of neutral
nitrogen and oxygen are similar, we assumed that the ionization fractions of
both elements are the same for singly ionized atoms, i.e.
N$^+/{\rm N}={\rm O}^+$/O (e.g. Sembach et al. 2000). 3) It has been observed in
several objects that the ratio of the oxygen abundance to the sulfur abundance
is about 100/3, independent of metallicity \citep{2allen}. We adopted this ratio
to derive the sulfur abundance from the oxygen abundance. 4) We assumed that
O$^{+++}$/O (and higher ionization stages of oxygen) is negligibly small and
therefore (O$^+/{\rm O})+({\rm O}^{++}/{\rm O})=1$. (No assumptions were made
for doubly ionized sulfur or doubly ionized nitrogen, i.e. we neither derived
nor used these ionization fractions.) 5) One additional assumption was
necessary, which yielded three different fitting methods. In Paper I, we
introduced these three methods. We adopted a constant ionization fraction
O$^{++}$/O along the slit (Method A), a constant temperature along the slit
(Method B), and a constant oxygen abundance along the slit (Method C). We
concluded that using a constant oxygen abundance (Method C) yielded the most
reasonable results in our simple models. In this paper, we will only use the
method of constant oxygen abundance for our analysis. Since it is important to
keep in mind how the properties were derived and what they depend on, we
summarize the important steps of our analysis here. A detailed description of
our method can be found in Paper I (Method C).

We combined the equations of the [\ion{O}{2}]/[\ion{O}{3}] and
[\ion{O}{3}]/H$\beta$ line ratios to eliminate the dependence on the ionization
fractions of oxygen (Paper I). [\ion{O}{3}]/H$\beta$ becomes
\begin{equation}
\frac{\rm [O\,III]}{\rm H\beta}=\frac{5.03\cdot 10^5\,T_4^{0.33}\,{\rm e}
^{-2.88/T_4}}{{\rm [O\,II]}/{\rm [O\,III]}\,0.40\,{\rm e}^{0.99/T_4}+1}\left(
\frac{O^{\rule{0cm}{1ex}}}{H}\right) \label{2e1}
\end{equation}
\citep{2oster}. All line ratio equations contain intensities in
ergs\,s$^{-1}$\,cm$^{-2}$\,sr$^{-1}$ with $T_4$ being the electron temperature
in 10\,000\,K. We assumed solar abundance for oxygen
(O/${\rm H}=8.5\cdot10^{-4}$ \citep{2dapp}). The assumption of solar oxygen
abundance yielded the most reasonable values for temperature, nitrogen abundance
and ionization fractions in Paper I (in comparison with the other methods
discussed in Paper I) with the indication that slight changes in the oxygen
abundance do not change the values of the derived properties significantly. We
therefore did not use this method with other values for the oxygen abundance.

We calculated the minimum and maximum allowed electron temperature for each data
point along the slit by fitting the observed [\ion{O}{3}]/H$\beta$ line ratios
within their 1\,$\sigma$ error bars using temperature increments/decrements of
100\,K in equation (\ref{2e1}). The allowable fractions of O$^{++}$ then were
found by fixing the [\ion{O}{3}]/H$\beta$ line ratio at its observed value, and
fitting this with our estimated maximum and minimum electron temperatures at
each point along the slit:
\begin{equation}
\frac{\rm [O\,III]}{\rm H\beta}=5.03\cdot10^5\,T_4^{0.33}\,{\rm
e}^{-2.88/T_4}\,\left(\frac{O^{++}}{O}\right)\,\left(\frac{O^{\rule{0cm}{1ex}}}
{H}\right)\,\left(\frac{H^+}{H}\right)^{-1}
\end{equation}
This approach is consistent with our parametric models where we seek to
demonstrate that variations in electron temperature can explain observed trends
in line ratios. However, it has the disadvantage of causing the predicted
[\ion{O}{3}]/H$\alpha$ ratios to coincide at the maximum and minimum electron
temperatures. In general, random errors are assumed to be dominated by the
uncertainty of the H$\beta$ line strength measurements, whereas errors in line
ratios of [\ion{O}{2}] or [\ion{O}{3}] over any red emission line (H$\alpha$,
[\ion{N}{2}], [\ion{S}{2}]) tend to be systematic and dominated by the
extinction correction.

For both calculated temperature extremes, we derived the nitrogen abundances and
the sulfur ionization fractions using the
equations
\begin{eqnarray}
{\rm [O\,II]/[N\,II]:}\quad\frac{I_{3727}}{I_{6583.4}} & = & 2.65\,{\rm e}^
{-1.69/T_4}\,\left(\frac{O^+}{O}\right)\,\left(\frac{O^{\rule{0cm}{1ex}}}
{H}\right)\,\left(\frac{N^+}{N}\right)^{-1}\,\left(\frac{N^{\rule{0cm}{1ex}}}
{H}\right)^{-1} \\
{\rm [O\,II]/[S\,II]:}\quad\frac{I_{3727}}{I_{6716.4}} & = & 0.58\,{\rm e}^
{-1.73/T_4}\,\left(\frac{O^+}{O}\right)\,\left(\frac{O^{\rule{0cm}{1ex}}}
{H}\right)\,\left(\frac{S^+}{S}\right)^{-1}\,\left(\frac{S^{\rule{0cm}{1ex}}}
{H}\right)^{-1}
\end{eqnarray}
\citep{2oster}. The final step was to compare the observed H$\alpha$ line ratios
with the predicted ratios based upon the derivations above:
\begin{eqnarray}
{\rm [O\,II]/H\alpha:}\quad\frac{I_{3727}}{I_{6562.8}} & = & 4.31\cdot10^5\,T_4
^{0.4}\,{\rm e}^{-3.87/T_4}\,\left(\frac{O^+}{O}\right)\,\left(\frac{O^{\rule
{0cm}{1ex}}}{H}\right)\,\left(\frac{H^+}{H}\right)^{-1} \\
{\rm [O\,III]/H\alpha:}\quad\frac{I_{5006.9}}{I_{6562.8}} & = & 1.74\cdot10^5\,
T_4^{0.4}\,{\rm e}^{-2.88/T_4}\,\left(\frac{O^{++}}{O}\right)\,\left(\frac{O^
{\rule{0cm}{1ex}}}{H}\right)\,\left(\frac{H^+}{H}\right)^{-1} \\
{\rm [N\,II]/H\alpha:}\quad\frac{I_{6583.4}}{I_{6562.8}} & = & 1.62\cdot10^5\,
T_4^{0.4}\,{\rm e}^{-2.18/T_4}\,\left(\frac{N^+}{N}\right)\,\left(\frac{N^{\rule
{0cm}{1ex}}}{H}\right)\,\left(\frac{H^+}{H}\right)^{-1} \\
{\rm [S\,II]/H\alpha:}\quad\frac{I_{6716.4}}{I_{6562.8}} & = & 7.49\cdot10^5\,
T_4^{0.4}\,{\rm e}^{-2.14/T_4}\,\left(\frac{S^+}{S}\right)\,\left(\frac{S^{\rule
{0cm}{1ex}}}{H}\right)\,\left(\frac{H^+}{H}\right)^{-1}
\end{eqnarray}
(Osterbrock 1989; Haffner et al. 1999).

\subsection{NGC\,5775}

\begin{figure}
\epsscale{0.8}
\plotone{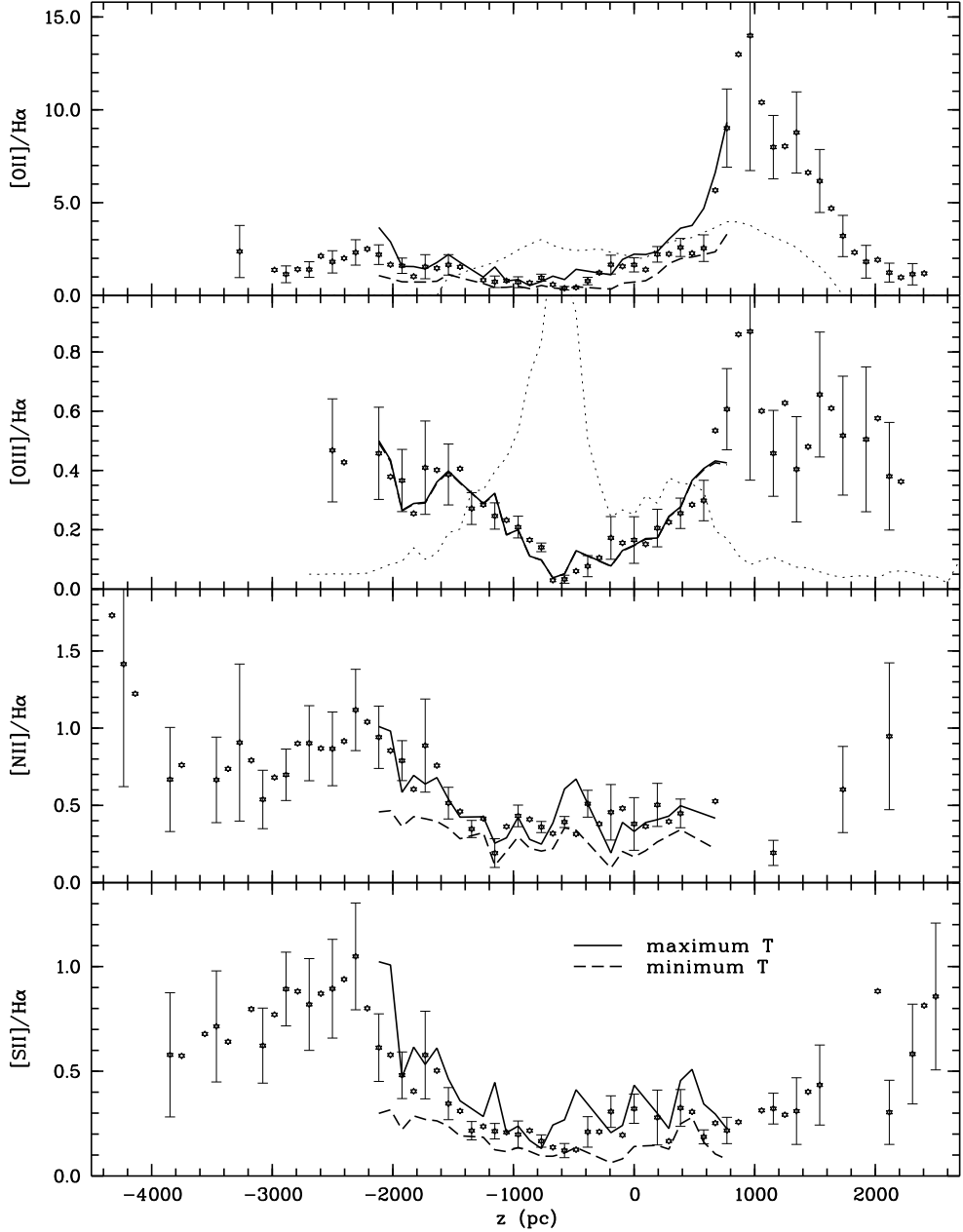}
\caption{\label{2f6}
H$\alpha$ line ratios and derived properties perpendicular to the disk of
NGC\,5775 (disk 2 slit). Note that the data have been smoothed by a running
average over 9 rows of spectrum (770\,pc, see Table \ref{2t2}), but we plot the
results for each row to show the full extent of these data. ({\em a}) The
measured H$\alpha$ line ratios are compared with the predicted line ratios using
the maximum temperature ({\em solid lines}) and the minimum temperature ({\em
dashed lines}). The {\em dotted line} in the {\em top panel} shows the derived
optical depth $\tau$ (unscaled). The {\em dotted line} in the {\em second from
top panel} shows the H$\alpha$ intensity along the slit scaled down to fit the
plot. The two predictions for [\ion{O}{3}]/H$\alpha$ are almost identical. Error
bars are plotted only for every other row of the spectrum. ({\em b}) The derived
electron temperature, ionization fraction of doubly ionized oxygen, the nitrogen
abundance, and the sulfur ionization fraction are shown for both the maximum
temperature ({\em filled circles}) and the minimum temperature ({\em open
diamonds}). The {\em dashed line} indicates solar nitrogen abundance. ({\em c})
same as ({\em a}), but without extinction correction. ({\em d}) same as ({\em
b}), but without extinction correction.}
\end{figure}

Figure \ref{2f6} shows the observed line ratios and the derived properties for
\objectname[]{NGC\,5775} (disk 2 slit) for both cases, extinction corrected and
uncorrected, as a function of distance from the midplane. Our disk slits cross
regions where extended emission from eDIG is seen in the deep
H$\alpha+$[\ion{N}{2}] images of \citet{2colea}. From these images we also
estimate that eDIG is likely to dominate at $z$ distances of $>1$\,kpc in this
slightly inclined galaxy. The optical depth as derived from H$\alpha$/H$\beta$
is relatively high ($\tau=3-4$) along the entire region in which we obtained
measurements of high enough signal--to--noise ratios. The applied extinction
correction yielded maximum factors of about 12 for [\ion{O}{2}], 3 for H$\beta$,
2.7 for [\ion{O}{3}] and about 1 for the red emission lines. Although the actual
numbers may differ a little between extinction corrected and uncorrected line
ratios, the increasing or decreasing trends in the intensity line ratios are the
same in both cases. The lowest line ratios correlate with the brightest line
emission.

The line ratio predictions using the maximum allowed temperature match the data
well within the error bars in the extinction corrected case (except for a small
``bump'' around $z=-500$\,pc). The predictions derived from the minimum allowed
temperature also match the data within the error bars or are somewhat lower for
[\ion{S}{2}]/H$\alpha$ and [\ion{N}{2}]/H$\alpha$. In the uncorrected case, the
maximum temperature yields too high values, whereas the fits of the minimum
temperature lie more or less within the error bars or within the $2\sigma$
range. The predictions derived from the maximum allowed temperature also match
the data reasonably, to within $\approx2\sigma$, with most of the problems
appearing in regions where extinction corrections are important.

The derived electron temperature increases with $|z|$ in both cases, although
the increase is stronger in the uncorrected case. The oxygen ionization fraction
decreases with $|z|$ on one side of the disk ($z<-1$\,kpc) with the uncorrected
values showing the stronger decrease, but increases on the other side
up to about 600\,pc, before it starts to decrease again. The nitrogen abundance
shows a large difference between the corrected and the uncorrected values. In
the extinction corrected case, N/H seems to decrease towards positive $z$ along
the slit, whereas the uncorrected N/H shows a maximum near the midplane (between
$-1$\,kpc and 0\,pc), but stays constant at solar value or slightly above along
the rest of the slit. The sulfur ionization fraction also shows a larger
difference between the two cases. The extinction corrected values decrease
towards positive $z$ along the slit, whereas the uncorrected values for S$^+$/S
decrease with $|z|$ on one side (negative $z$), but only slightly decrease or
stay constant with increasing $z$ on the other side. Most values for S$^+$/S
derived from the maximum allowed temperature are greater than unity and thus not
physical. Therefore, the maximum allowed temperature can be ruled out as an
option in this case. The other slit (disk 1) shows a second peak in H$\alpha$ at
$z\approx +800$\,pc, thus possibly associated with ionized gas in the disk. The
[\ion{O}{2}]/H$\alpha$ line ratio reaches a local minimum in that area. However,
since the other emission lines have too low signal--to--noise ratios, we could
not analyze them and therefore do not discuss them here.

\subsection{NGC\,3079}

\begin{figure}
\plotone{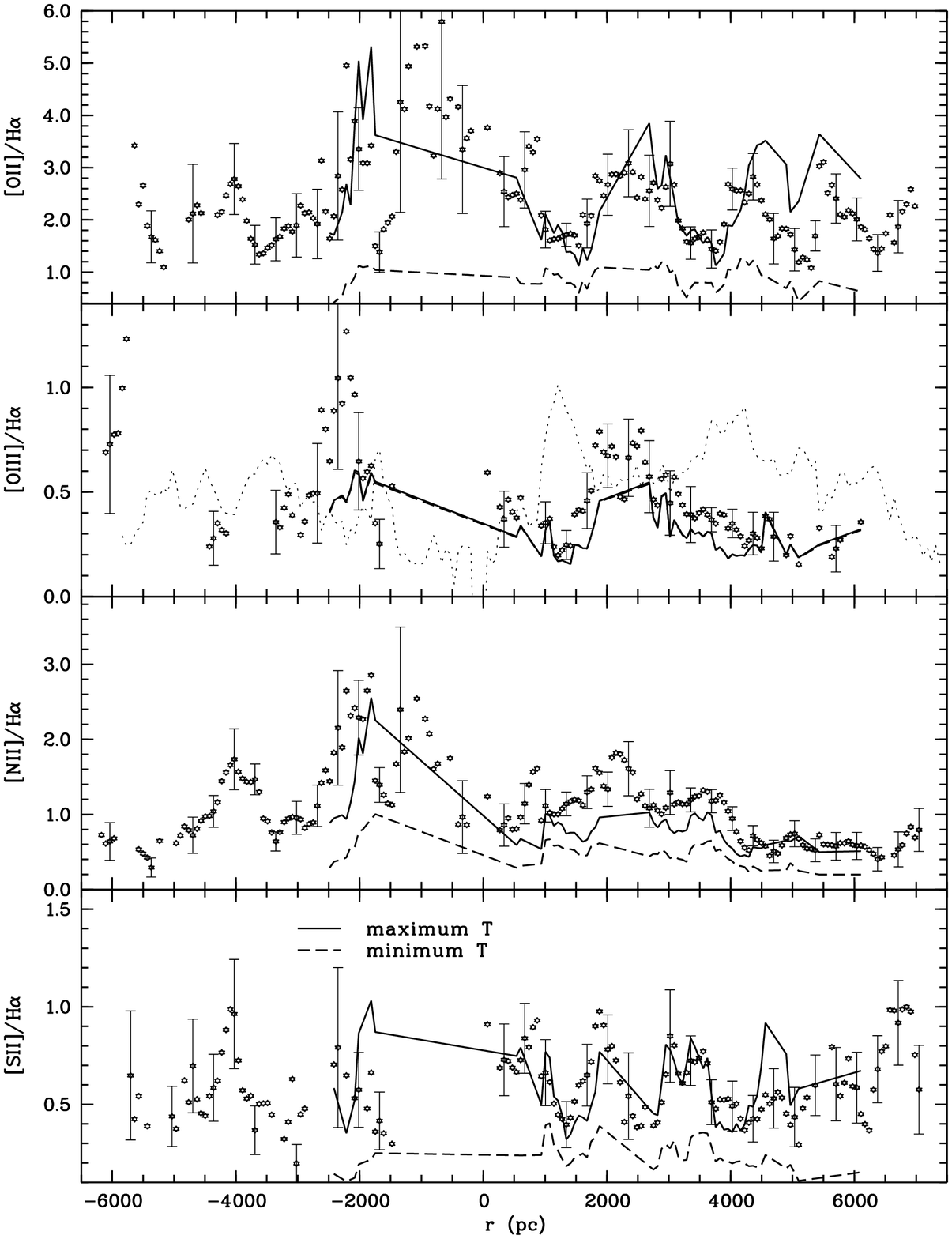}
\caption{\label{2f7}
H$\alpha$ line ratios and derived properties parallel to the disk of NGC\,3079.
Note that the data have been smoothed by a running average over 5 rows of
spectrum (270\,pc, see Table \ref{2t2}), but we plot the results for each row to
show the full extent of these data. ({\em a}) The measured H$\alpha$ line ratios
are compared with the predicted line ratios using the maximum temperature ({\em
solid lines}) and the minimum temperature ({\em dashed lines}). The {\em dotted
line} in the {\em second from top panel} shows the H$\alpha$ intensity along the
slit scaled down to fit the plot. The two predictions for [\ion{O}{3}]/H$\alpha$
are almost identical. Error bars are plotted only for every 5th row of the
spectrum. ({\em b}) The derived electron temperature, ionization fraction of
doubly ionized oxygen, the nitrogen abundance, and the sulfur ionization
fraction are shown for both the maximum temperature ({\em filled circles}) and
the minimum temperature ({\em open diamonds}). The {\em dashed line} indicates
solar nitrogen abundance.}
\end{figure}

Figure \ref{2f7} shows the derived electron temperatures, abundances and
ionization fractions as well as the predicted line ratios for the DIG slit of
\objectname[]{NGC\,3079} at a constant distance above the midplane. The observed
[\ion{O}{2}]/H$\alpha$ line ratio shows variations beyond $r=+3$\,kpc which are
not present in the other line ratios. The line ratio predictions using the
minimum allowed temperature are too low in comparison with the observed line
ratios. The line ratio predictions derived from the maximum allowed temperature
are a little low for the [\ion{N}{2}]/H$\alpha$ and [\ion{O}{3}]/H$\alpha$ line
ratios and somewhat higher for [\ion{O}{2}]/H$\alpha$ at about $z=+4.5$\,kpc,
but in general, the predictions match the observed data within the error bars.
The derived electron temperature varies along the slit by about 1500--2000\,K.
It is lower in regions of stronger H$\alpha$ emission, i.e. at
$r\approx +1500$\,pc and $+3700$\,pc (we will discuss this in more detail in
section \ref{2s4}); however, there is no systematic trend with increasing
distance from the estimated center of the galaxy. The oxygen ionization fraction
stays relatively constant with a few variations along the slit. The nitrogen
abundance seems to decrease slightly with increasing $|r|$, as expected for a
radial chemical abundance gradient. In Figure \ref{2f7}b we also find an
apparent rapid increase in nitrogen abundance towards the end of our slit, at
$r\approx-2$\,kpc. This suggests that a problem exists either with the model or
the data, and in either case, the approach adopted here probably is not yielding
reliable results. The sulfur ionization fraction appears to stay more or less
constant along the slit.

\subsection{NGC\,4631}

\begin{figure}
\plotone{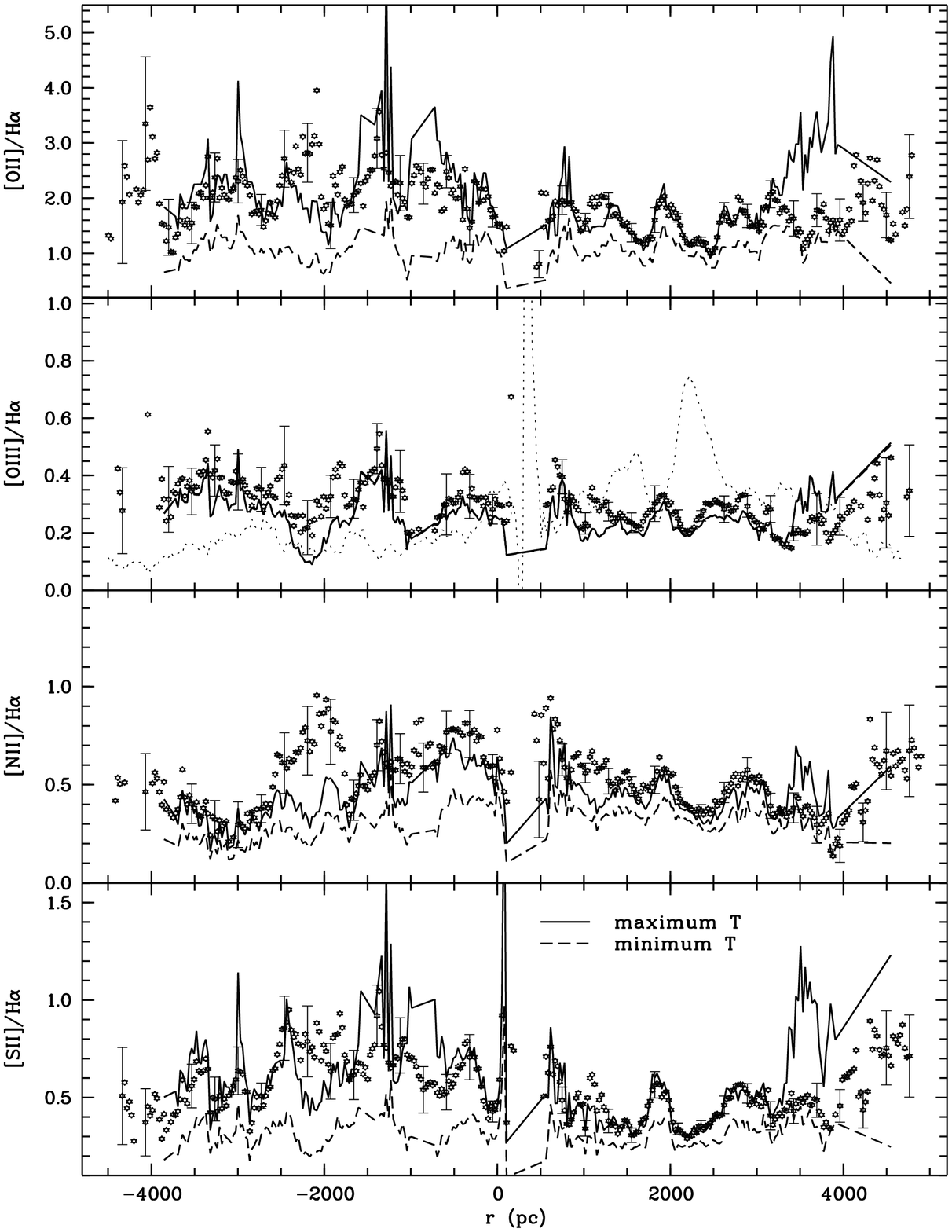}
\caption{\label{2f8}
Same as Fig. \ref{2f7}, but for NGC\,4631 (DIG 2 slit). Note that the data have
been smoothed by a running average over 9 rows of spectrum (215\,pc, see Table
\ref{2t2}), but we plot the results for each row to show the full extent of
these data. Error bars are plotted only for every 10th row of the spectrum.}
\end{figure}

The derived properties and line ratios for the DIG 2 slit (parallel to the disk)
of \objectname[]{NGC\,4631} are shown in Figure \ref{2f8}. All line ratios are
higher in the western part of the slit (negative $r$) than in the eastern part.
Variations in the line ratios coincide with variations in the emission line
intensities with brighter regions having lower line ratios (cf Martin \& Kern
2001). The [\ion{O}{2}]/H$\alpha$ and [\ion{O}{3}]/H$\alpha$ line ratios behave
similarly except for two regions. At $r\approx -2200$\,pc,
[\ion{O}{2}]/H$\alpha$ reaches a maximum, whereas [\ion{O}{3}]/H$\alpha$ shows a
minimum. Only a part of the brighter emission region at
$+2000\,{\rm pc}<r<+2500$\,pc has lower [\ion{O}{3}]/H$\alpha$ values than the
surrounding area. The other three line ratios have minima in this region. All
line ratios appear to slightly decrease with $r$ or to stay constant in the
eastern part of the slit up to about $r=+4$\,kpc, where the line ratios seem to
increase with increasing distance from the center. The line ratios appear to
stay constant in the western part of the slit ($r<0$) with a possible decrease
in [\ion{S}{2}]/H$\alpha$ and [\ion{N}{2}]/H$\alpha$ beyond $r=-2$\,kpc.

The line ratio predictions using the minimum allowed electron temperature are
too low along the entire slit, as it is the case in \objectname[]{NGC\,3079}
described above. The maximum temperature predictions match the observed data
within the error bars except for the [\ion{O}{2}]/H$\alpha$ line ratio at
$r>+3.5$\,kpc and $r\approx-800$\,pc and to a lesser extent also the
[\ion{S}{2}]/H$\alpha$ ratio in the same areas. The H$\alpha$/H$\beta$ line
ratio is higher in these regions ($\approx 4.5$) implying that extinction
correction is necessary in these areas. [\ion{O}{2}]/H$\alpha$,
[\ion{N}{2}]/H$\alpha$, and [\ion{S}{2}]/H$\alpha$ show a mismatch between
observed and predicted line ratios at $r\approx-2.2$\,kpc which is also related
to variations in the H$\alpha$/H$\beta$ line ratio. We did not try to correct
for extinction, because the H$\beta$ emission line is weak in these areas. The
derived electron temperature seems to stay constant for most of the slit with
somewhat higher values at $r<0$ than at $r>0$, but the temperature appears to
increase with $r$ beyond $r\approx +3$\,kpc. The oxygen ionization fraction, on
the other hand, stays more or less constant at $r>0$, but slightly increases
with increasing $|r|$ at $r<-3$\,kpc. The nitrogen abundance decreases slightly
with $|r|$ on both sides of the slit and is lower than the solar abundance along
the entire slit. N/H shows a minimum at $r=-3.2$\,kpc. It is possible that the
nitrogen ionization fraction decreases in this area instead of the nitrogen
abundance, that is the second of our basic assumptions (N$^+/{\rm N}={\rm
O}^+$/O, cf section \ref{2s3}) could be inappropriate in this area. The sulfur
ionization fraction behaves similarly to the electron temperature, i.e. it stays
constant with higher values in the western part ($r<0$) than in the eastern
part, but increases with $r$ in the eastern part beyond $r\approx +2.5$\,kpc.

The emission in the other parallel slit (called DIG 1), which is at larger
distance from the midplane, is very faint and did not yield enough reliable
measurements. Therefore, we do not show the line ratios here, but briefly want
to mention that [\ion{O}{2}]/H$\alpha$ and [\ion{O}{3}]/H$\alpha$ values could
be obtained in the area $+1\,{\rm kpc}<r<+3$\,kpc. The average values
($1.8\pm0.2$ for [\ion{O}{2}]/H$\alpha$ and $0.45\pm0.05$ for
[\ion{O}{3}]/H$\alpha$) are a little higher than those of the DIG 2 slit in this
area ($1.6\pm0.2$ and $0.27\pm0.03$, respectively). The average electron
temperature of the DIG 1 slit is $T_4=0.72\pm0.06$ at $0<r<+3$\,kpc, while the
variations in temperature are not correlated with those observed in the DIG 2
slit shown in Figure \ref{2f8}b. The average oxygen ionization fraction of the
DIG 1 slit in the same area is $0.12\pm0.03$.

\subsection{NGC\,4634}

\begin{figure}
\plotone{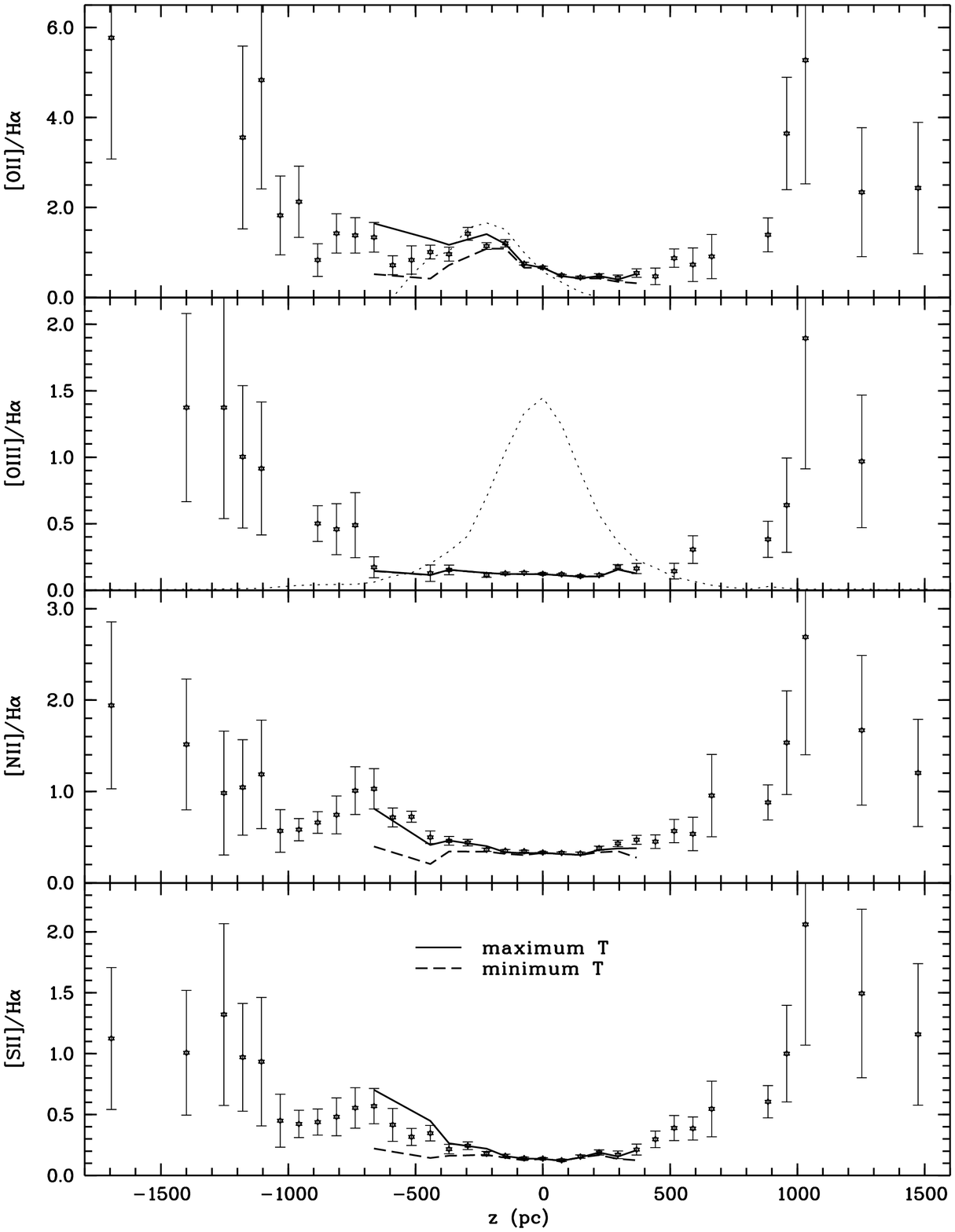}
\caption{\label{2f9}
Same as Fig. \ref{2f6}, but for NGC\,4634 (disk slit). Note that the data have
been smoothed by a running average over 9 rows of spectrum (590\,pc, see Table
\ref{2t2}) at $|z|>700$\,pc, but we plot the results for each row to show the
full extent of these data.}
\end{figure}

For \objectname[]{NGC\,4634}, the line ratios and derived properties of the disk
slit (i.e. perpendicular to the plane) are shown in Figure \ref{2f9}. The
derived optical depth $\tau$ is relatively low in comparison with the optical
depth in other galaxies (see Paper I for comparison). The applied extinction
correction yielded a maximum factor of only 2.9 for [\ion{O}{2}] (unlike 9--12
in Paper I), 1.6 for H$\beta$ and 1.5 for [\ion{O}{3}], whereas the red emission
lines had factors of 1. Figure \ref{2f9} shows both the extinction corrected
line ratios and properties and the uncorrected values for each.

Like in the galaxies discussed above, the line ratios in
\objectname[]{NGC\,4634} are lowest (even [\ion{O}{3}]/H$\alpha$), where the
emission lines are brightest. The peak in the [\ion{O}{2}]/H$\alpha$ line ratio
at $z\approx -250$\,pc (Fig. \ref{2f9}a) is introduced by the extinction
correction. The predicted line ratios in the extinction corrected case are again
too low for the minimum allowed temperature (as observed in the other galaxies).
The predictions using the maximum temperature match the observed data reasonably
well ([\ion{O}{3}]/H$\alpha$, [\ion{N}{2}]/H$\alpha$) or are slightly higher
than the observed data ([\ion{O}{2}]/H$\alpha$ and [\ion{S}{2}]/H$\alpha$ at
$r\approx -500$\,pc). The uncorrected line ratio predictions are too high in the
area where $\tau>0$ ($-600\,{\rm pc}<z<+200$\,pc).

The maximum allowed electron temperature increases with increasing distance from
the midplane in the southwestern part of the slit ($z<0$) in both cases,
extinction corrected as well as uncorrected derivations. The minimum allowed
temperature first increases with increasing height and then decreases at about
$z=-250$\,pc. The oxygen ionization fraction slightly decreases towards lower
$z$ (for $T=T_{\rm max}$) or maybe stays constant (for $T\approx T_{\rm min}$).
The trend in the nitrogen abundance, however, depends somewhat on the extinction
correction, because it was derived using blue and red lines. The extinction
corrected N/H shows a minimum at about $z=-200$\,pc, whereas the uncorrected
abundance shows a minimum near $z=0$. Both suggest that N/H increases with
$|z|$. We observed this strong variation between the two cases in other galaxies
as well (see Paper I). The sulfur ionization fraction increases with $|z|$ in
both cases, following the same shift of the minimum as the nitrogen abundance.

\begin{figure}
\plotone{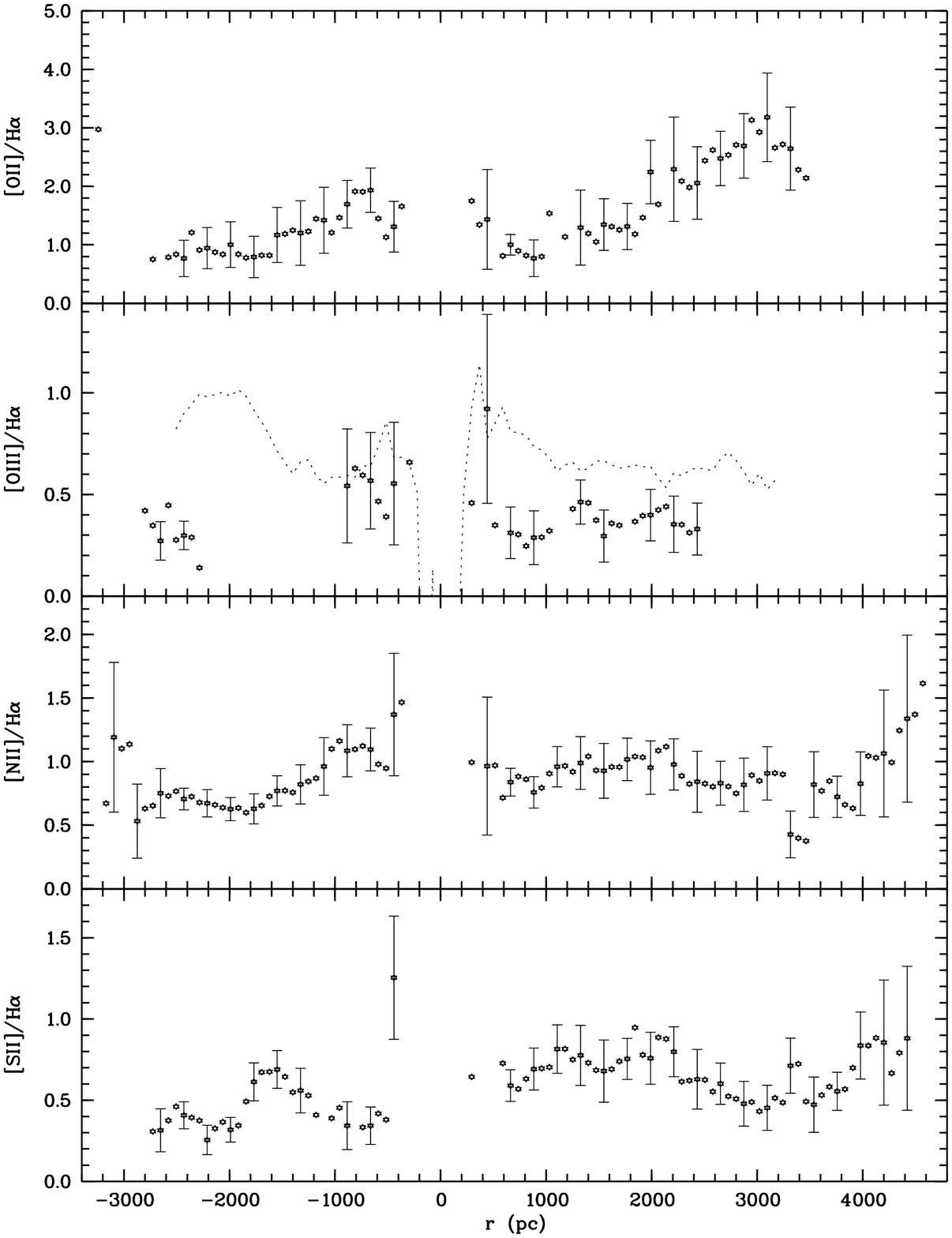}
\caption{\label{2f10}
H$\alpha$ line ratios parallel to the disk of NGC\,4634 (DIG slit). Note that
the data have been smoothed by a running average over 9 rows of spectrum
(590\,pc, see Table \ref{2t2}), but we plot the results for each row to show the
full extent of these data. The {\em dotted line} in the {\em second from top
panel} shows the H$\alpha$ intensity along the slit scaled down to fit the plot.
Error bars are plotted only for every 3rd row of the spectrum.}
\end{figure}

The line ratios of the slit parallel to the disk of \objectname[]{NGC\,4634}
(DIG slit) are shown in Figure \ref{2f10}. Unfortunately, the H$\beta$ emission
line was too weak to yield reliable measurements. Extinction correction did not
seem necessary. The [\ion{O}{2}]/H$\alpha$ line ratio appears to overall
decrease towards lower $r$, while the emission line intensities increase towards
lower $r$. Around $r=-600$\,pc, the line ratios [\ion{N}{2}]/H$\alpha$ and
[\ion{O}{3}]/H$\alpha$ are both slightly higher than at the rest of the slit.
[\ion{N}{2}]/H$\alpha$ appears to increase beyond $r=+4$\,kpc and basically
stays constant near the center of the galaxy. However, this line ratio shows two
broad minima at $-3\,{\rm kpc}<r<-1$\,kpc and $+2.2\,{\rm kpc}<r<+4$\,kpc. Both
minima seem to exist also in the [\ion{S}{2}]/H$\alpha$ line ratio. Although we
were not able to apply our analysis method to this slit position, the observed
line ratios might still be useful in comparison with the other DIG slits. For
example, a comparison with the H$\alpha$ line ratios and the emission line
intensity of the DIG slit of \objectname[]{NGC\,4631} suggests that the minimum
around $r=-2$\,kpc in [\ion{N}{2}]/H$\alpha$ and [\ion{S}{2}]/H$\alpha$ is
caused by temperature variations, whereas the minimum around $r=+3$\,kpc is
caused by decreases in abundance or ionization fraction of nitrogen and sulfur.

\subsection{NGC\,891}

\begin{figure}
\plotone{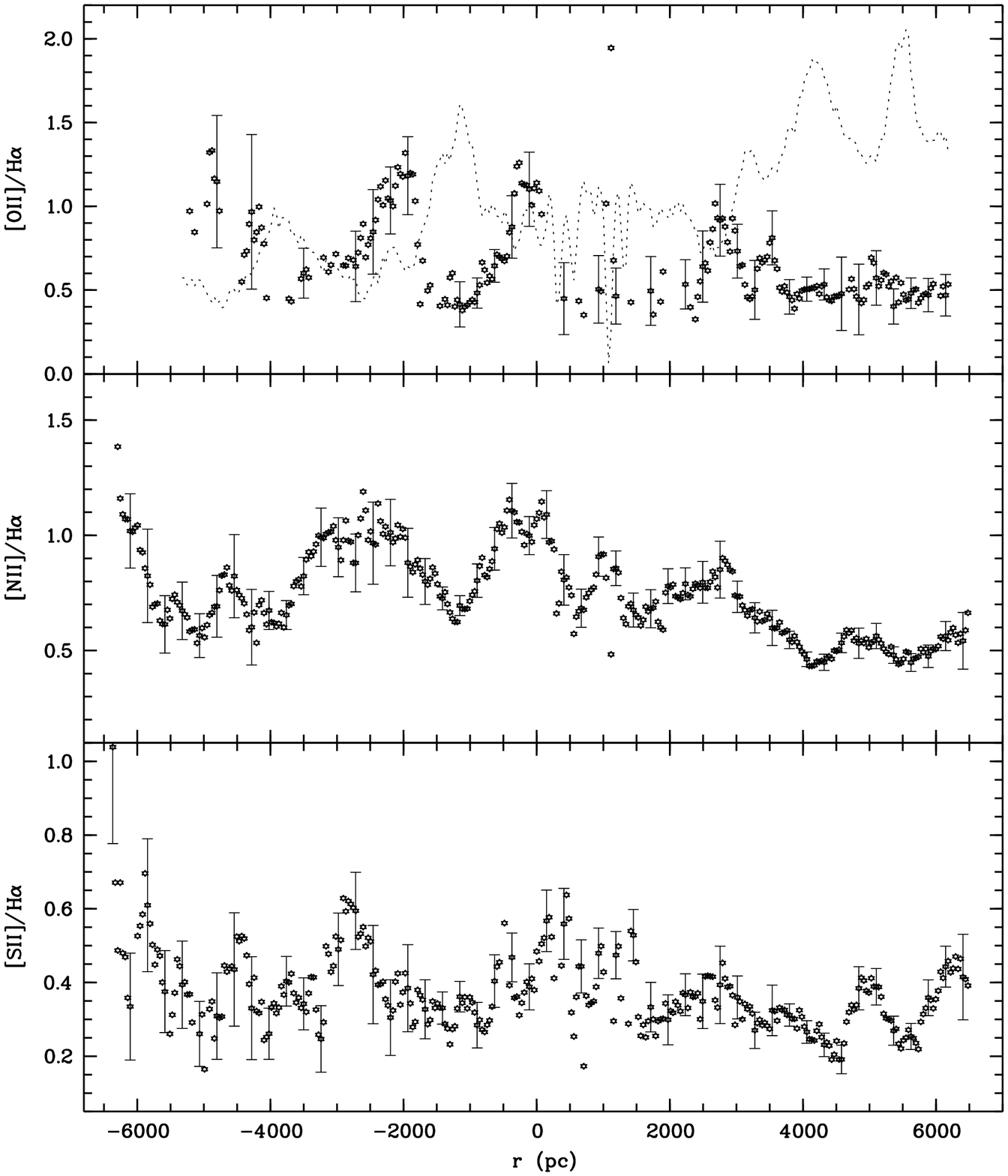}
\caption{\label{2f11}
H$\alpha$ line ratios parallel to the disk of NGC\,891. Note that the data have
been smoothed by a running average over 9 rows of spectrum (300\,pc, see Table
\ref{2t2}), but we plot the results for each row to show the full extent of
these data. The {\em dotted line} in the {\em top panel} shows the H$\alpha$
intensity along the slit scaled down to fit the plot. Error bars are plotted
only for every 7th row of the spectrum.}
\end{figure}

We were not able to obtain good measurements of the H$\beta$ line for
\objectname[]{NGC\,891} (DIG slit). We therefore could not derive the optical
depth for this slit position. However, we do not expect the extinction to be
significant in these observations. We were also not able to measure
[\ion{O}{3}], leaving us only with the line ratios [\ion{O}{2}]/H$\alpha$,
[\ion{N}{2}]/H$\alpha$, and [\ion{S}{2}]/H$\alpha$, which are shown in Figure
\ref{2f11}. The line ratios are on average higher and show stronger variations
in the southwestern part of the slit ($r<0$) than in the northeastern part. The
variations in the line ratios seem to correlate again with the brightness of the
emission lines, i.e. brighter line emission yields lower line ratios. However,
even though the variations in the line emission have more or less the same
magnitude on both sides of the slit, the variations in the line ratios differ
between the two sides (0.2--0.3 at $r>0$ and $\geq 0.6$ at $r<0$ for both
[\ion{O}{2}]/H$\alpha$ and [\ion{N}{2}]/H$\alpha$).

\section{\label{2s4} DISCUSSION}

In general all our galaxies follow the trend of higher forbidden line
intensities relative to H$\alpha$ in regions of lower line emission. As a
consequence, the line ratios of slit positions cutting through the disk increase
with increasing $|z|$. However, there are exceptions: [\ion{O}{3}]/H$\alpha$ and
[\ion{O}{2}]/H$\alpha$ at $z>+1$\,kpc in \objectname[]{NGC\,5775} (disk 2) and
apparently all H$\alpha$ line ratios at $z>+1$\,kpc in \objectname[]{NGC\,4634}.
The line ratios of the slits parallel to the midplane show strong small scale
variations probably due to structures in the DIG. The line ratios have on
average lower values near overall brighter emission regions. This behavior has
been observed in other galaxies as well including both irregulars and spirals
and covering a range in properties (e.g. Ferguson, Wyse, \& Gallagher 1996;
Domg\"orgen \& Dettmar 1997; Otte \& Dettmar 1999; Haffner et al. 1999).

Over the last year, two other groups have obtained spectroscopic data of
\objectname[]{NGC\,5775}, \citet{2tuell} (hereafter TDSUR) and \citet{2col}
(hereafter CR). TDSUR's slit position is the same as our disk 2 slit, whereas
CR's slit 2 lies a little more towards the center of
\objectname[]{NGC\,5775}. We assumed the same distance to
\objectname[]{NGC\,5775} as CR (24.8\,Mpc) and consider in our comparisons the
slightly different distance used by TDSUR (26.7\,Mpc). Our
[\ion{N}{2}]/H$\alpha$ and [\ion{S}{2}]/H$\alpha$ measurements match those
observed by TDSUR and CR within the uncertainties except for the minimum beyond
$r=-2$\,kpc in our data, which is possibly observed to a lesser extent by TDSUR,
but not existent in CR's data. CR's [\ion{N}{2}]/H$\alpha$ measurements are a
little lower than TDSUR's and ours (0.55 as opposed to 0.7 or 0.8 at $z=-2$\,kpc
(SW)). The difference in the line ratios between TDSUR's and our measurements
increases towards bluer lines. CR's and our [\ion{O}{3}]/H$\alpha$ values are
higher than those measured by TDSUR (about 0.4 at $z=-2$\,kpc (SW) as opposed to
0.2, and about 0.1-0.15 at 0\,kpc as opposed to $<0.05$). The difference between
TDSUR's and our measurements is largest in the [\ion{O}{2}]/H$\alpha$ values.
TDSUR measure values from 0.05 to 0.4 between 0\,kpc and $-2$\,kpc, whereas our
measurements yield values of 0.4 (uncorrected) at 0\,kpc and 1.6 at $-2$\,kpc,
where extinction correction was no longer necessary. Unfortunately, CR did not
observe the [\ion{O}{2}] emission line and therefore cannot shed light on this
large discrepancy in the [\ion{O}{2}]/H$\alpha$ line ratios. Despite this large
difference in the [\ion{O}{2}]/H$\alpha$ values, the derived electron
temperatures and temperature variations are comparable between TDSUR, who used
the [\ion{N}{2}]/H$\alpha$ ratio to derive $T_4$, and our measurements using the
[\ion{O}{2}]/H$\alpha$ ratio. The temperatures are also in agreement with CR's
temperature values (also derived from the [\ion{N}{2}]/H$\alpha$ ratio). Our
electron temperature increases from $0.65\pm0.05$ (uncorrected) or $0.7\pm0.05$
(extinction corrected) at 0\,kpc to $0.75\pm0.07$ at $-2$\,kpc (using the
average between minimum and maximum allowed temperature), while TDSUR and CR
measure $T_4\approx 0.67$ at 0\,kpc and $T_4\approx 0.78$ at $-2$\,kpc (SW).
This suggests that [\ion{O}{2}]/H$\alpha$ and [\ion{N}{2}]/H$\alpha$ may be
equally reliable in deducing variations in temperature in the DIG.

\citet{2td} observed \objectname[]{NGC\,4634} as well. Their slit position s1 is
the same as our disk slit of \objectname[]{NGC\,4634}. Our
[\ion{N}{2}]/H$\alpha$ and [\ion{S}{2}]/H$\alpha$ line ratios match their
measurements within the error bars near the midplane. However, at larger
distances from the disk T\"ullmann \& Dettmar measured lower line ratios than we
did. In particular, the minimum at about $z=-1$\,kpc and the decrease beyond
$z=+1$\,kpc do not exist in their data. Unlike T\"ullmann \& Dettmar, we could
obtain [\ion{O}{2}] and [\ion{O}{3}] measurements. (Both groups also observed
\objectname[]{NGC\,4302} and could not get reliable measurements of this galaxy,
because the emission lines were too faint, whereas CR could at least measure the
red emission lines.)

The oxygen and sulfur ionization fractions derived from our observations stay
more or less constant in the DIG slits parallel to the midplane of our galaxies.
The behavior of O$^{++}$/O in the slits cutting through the disk is less clear
due to the variations in O$^{++}$/O along the slit and the uncertainty of the
extinction correction (although the dependence of O$^{++}$/O on the applied
extinction correction is less than in the galaxies discussed in Paper I). The
behavior of S$^+$/S also depends strongly on the extinction correction. The
values of the ionization fractions of sulfur and oxygen (including the
measurements of \objectname[]{NGC\,891}, \objectname[]{NGC\,4631}, and
\objectname[]{NGC\,3079} in Paper I) suggest that the ionization of the diffuse
gas likely varies with distance from the midplane. However, it does not seem to
vary on a large scale along the miplane at a constant height. If not all of the
gas is ionized (for example H$^+/{\rm H}=0.75$), the total amount of ionized
oxygen should be reduced by a similar factor (in this example (O$^+/{\rm
O})+({\rm O}^{++}/{\rm O})=0.75$) due to the similar ionization potentials.
Keeping the other basic assumptions unchanged (see beginning of section
\ref{2s3}), the derived electron temperature and nitrogen abundance do not
change in comparison with the H$^+/{\rm H}=1$ model, because the factor cancels
out. However, the derived ionization fractions for oxygen and singly ionized
sulfur are reduced by the same factor as the hydrogen ionization fraction. Thus,
rather constant O$^{++}$/O and S$^+$/S in the H$^+/{\rm H}=1$ model turn to
decreasing trends with increasing $|r|$ or $|z|$ when a decrease in H$^+$/H is
assumed with increasing distance from the galactic center or disk.

Smaller scale variations in the ionization fractions seem to be real and may be
caused by different features or density fluctuations in the ionized gas as these
variations are often related to emission line peaks. This is true for the
smaller scale variations with amplitudes higher than the scatter in all the
derived properties and line ratios (except maybe for peaks that were introduced
or enhanced by the extinction correction). The extent of these variations is
larger than the averaging done before the line measurements were taken, i.e. the
variations cannot have been introduced by our averaging. \citet{2make} also
observed fluctuations in the line ratios of \objectname[]{NGC\,4631} and
concluded that they are probably the result of multiple phases in the
interstellar medium.

Since the values of the nitrogen abundance also depend strongly on the
extinction correction, it is not clear whether N/H decreases with increasing
distance from the midplane in \objectname[]{NGC\,4634} and
\objectname[]{NGC\,5775}. Our DIG slits suggest that N/H decreases with
increasing distance from the center, i.e. that the abundance is lower farther
out in the galactic disks, where less star formation occurs. Since nitrogen is a
secondary product of nucleosynthesis, its abundance depends more strongly on
star formation and star formation history than the sulfur or oxygen abundances
(which we assumed to be constant along the slits) and therefore is expected to
be lower in the outer regions of the disks of galaxies. A decrease in oxygen
abundance towards the outer regions of the galactic disks causes our models to
predict an increase in electron temperature and to enhance the observed decrease
in nitrogen abundance. In the DIG slit of \objectname[]{NGC\,4631}, for example,
a decrease in oxygen abundance from solar at 0\,kpc to half the solar value at
about $|r|=3$\,kpc (i.e. decrease by a factor of 2) in the context of our models
leads us to find a decrease in N/H from $9\cdot10^{-5}$ at $r=0$\,kpc to
$4\cdot10^{-5}$ at $|r|=3$\,kpc, i.e. a decrease which equals or exceeds that of
O/H. (In the individual models of constant oxygen abundance, N/H decreases by a
factor of $\approx1.5$ from 0 to 3\,kpc.) The temperature increases by about
1000\,K at $|r|=3$\,kpc relative to the value obtained from the solar O/H model
at the same position, yielding an overall increase of $T_4$ with $|r|$. The
trends in the ionization fractions are less obvious. S$^+$/S increases by about
0.15, and O$^{++}$/O increases by about $0.01-0.02$ at $|r|=3$\,kpc when
compared to the values from the solar O/H model at that position resulting in a
possible increase in S$^+$/S and O$^{++}$/O with $|r|$ in
\objectname[]{NGC\,4631}. However, we have to keep in mind that changing other
basic assumptions listed at the beginning of section \ref{2s3} would change the
derived properties. For example, variations in nitrogen abundance could become
smaller, if we allowed changes in the nitrogen ionization fraction independently
from those of O$^+$/O.

\begin{figure}
\plotone{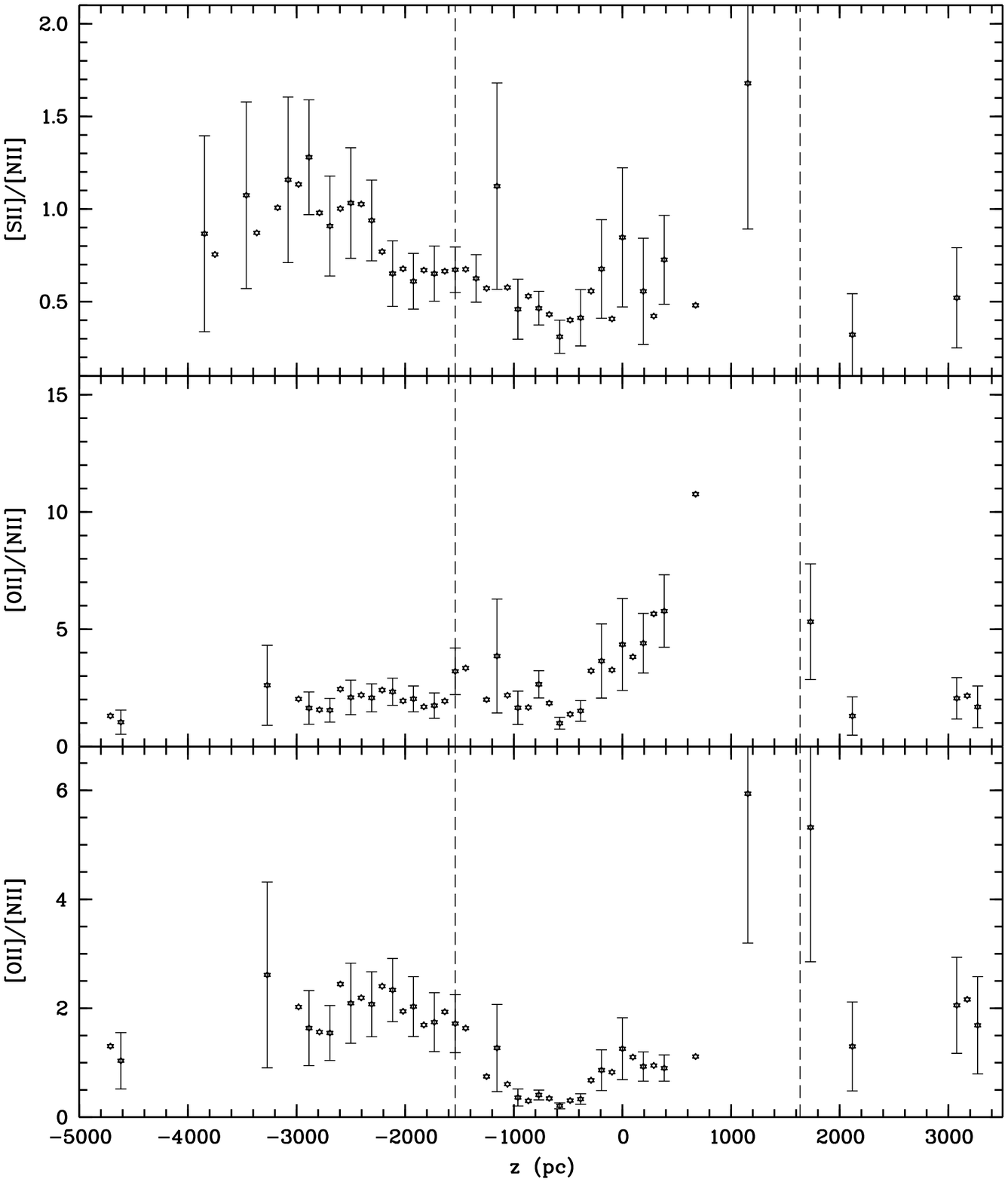}
\caption{\label{2f12}
[\ion{S}{2}]/[\ion{N}{2}] and [\ion{O}{2}]/[\ion{N}{2}] line ratios for
NGC\,5775 (disk 2 slit). Note that the data have been smoothed by a running
average over 9 rows of spectrum (770\,pc, see Table \ref{2t2}), but we plot the
results for each row to show the full extent of these data. {\em Top panel}:
[\ion{S}{2}]/[\ion{N}{2}] (extinction corrected). {\em Middle panel}:
[\ion{O}{2}]/[\ion{N}{2}] (extinction corrected). {\em Bottom panel}:
[\ion{O}{2}]/[\ion{N}{2}] (without extinction correction). The {\em dashed
lines} show the range affected by the extinction correction. Error bars are
plotted only for every other row of the spectrum.}
\end{figure}

\begin{figure}
\epsscale{1.0}
\plotone{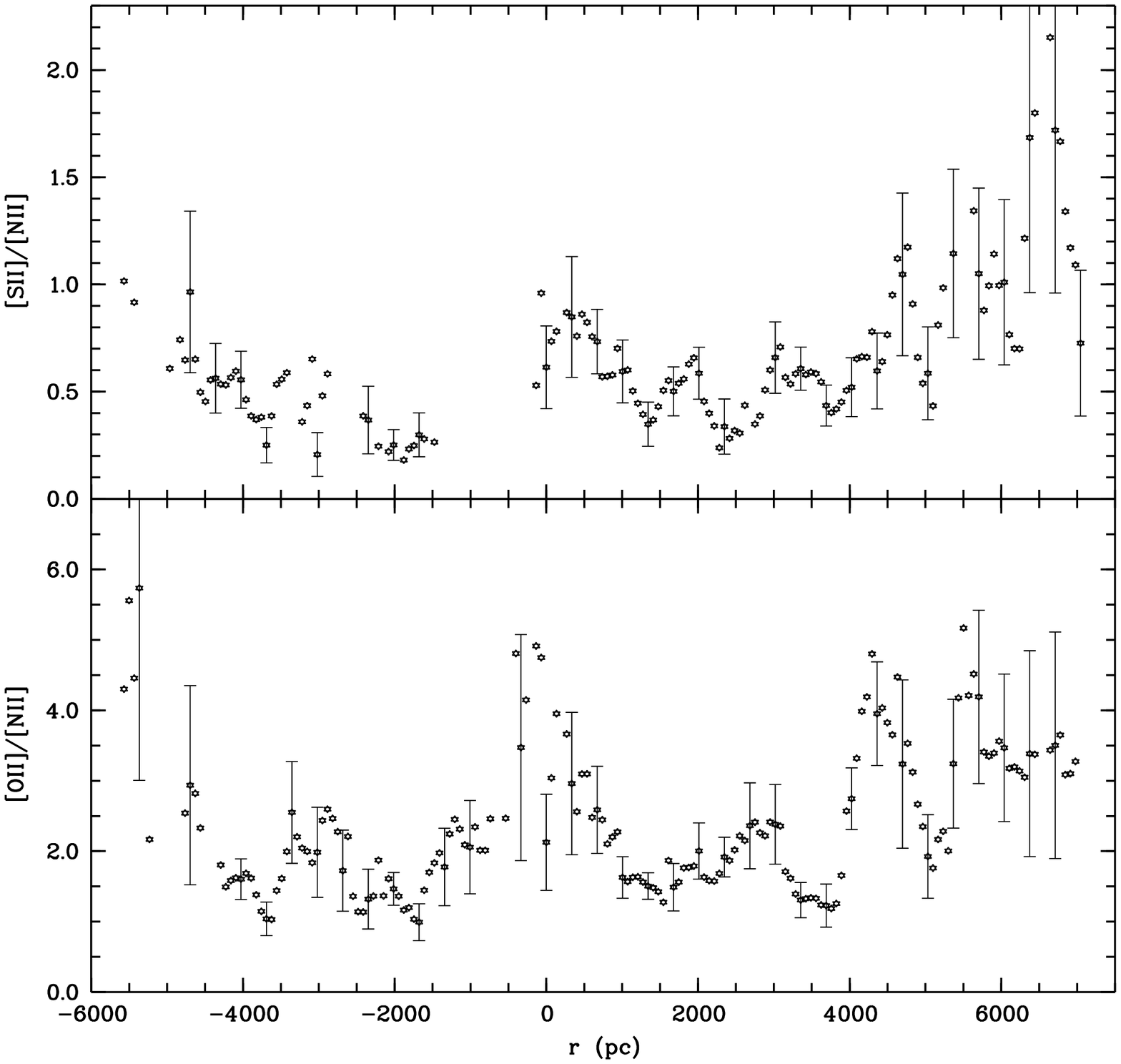}
\caption{\label{2f13}
[\ion{S}{2}]/[\ion{N}{2}] and [\ion{O}{2}]/[\ion{N}{2}] line ratios for
NGC\,3079. Note that the data have been smoothed by a running average over 5
rows of spectrum (270\,pc, see Table \ref{2t2}), but we plot the results for
each row to show the full extent of these data. Error bars are plotted only for
every 5th row of the spectrum.}
\end{figure}

\begin{figure}
\plotone{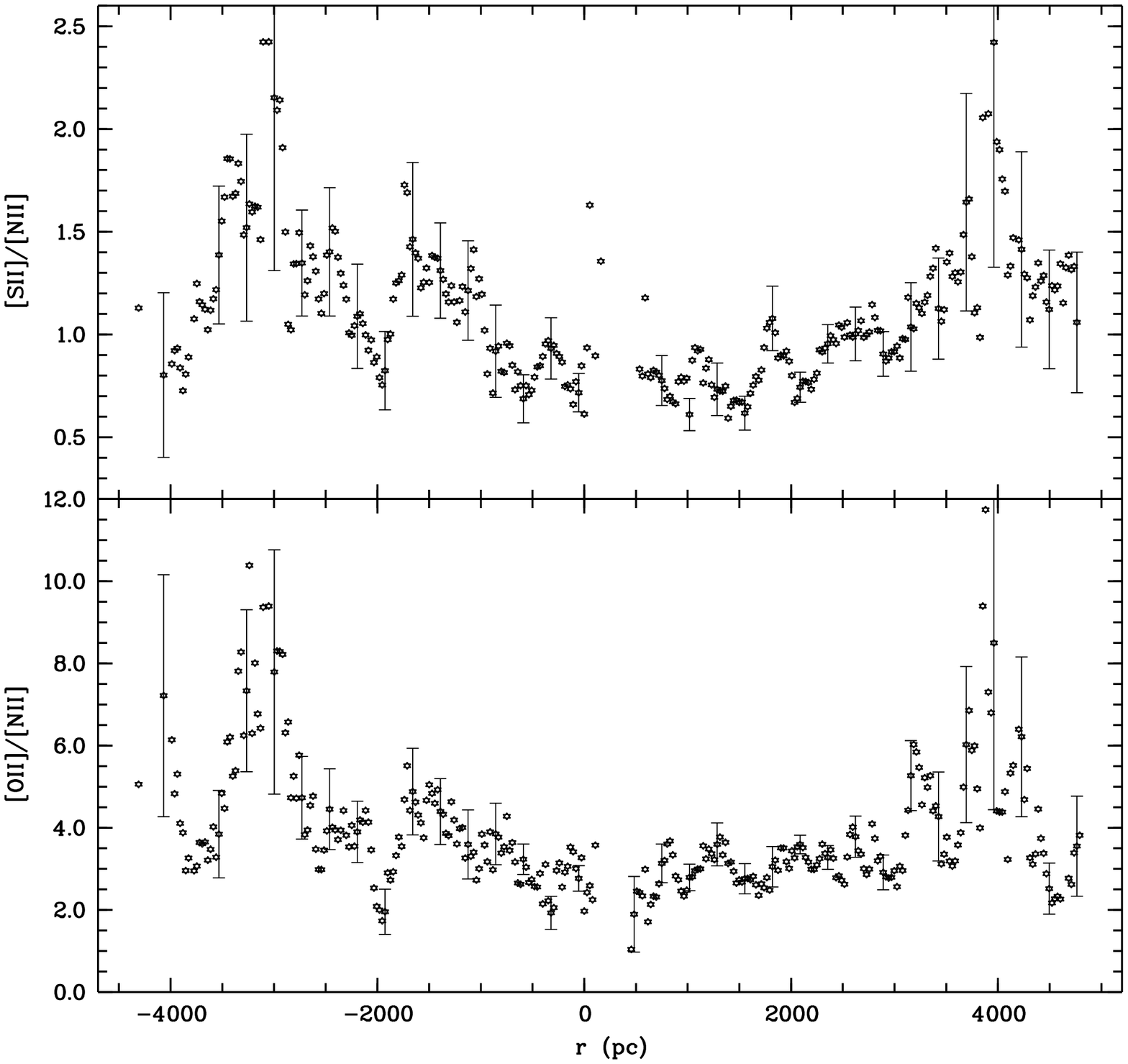}
\caption{\label{2f14}
Same as Fig. \ref{2f13}, but for NGC\,4631 (DIG 2 slit). Note that the data have
been smoothed by a running average over 9 rows of spectrum (215\,pc, see Table
\ref{2t2}), but we plot the results for each row to show the full extent of
these data. Error bars are plotted only for every 10th row of the spectrum.}
\end{figure}

\begin{figure}
\epsscale{0.8}
\plotone{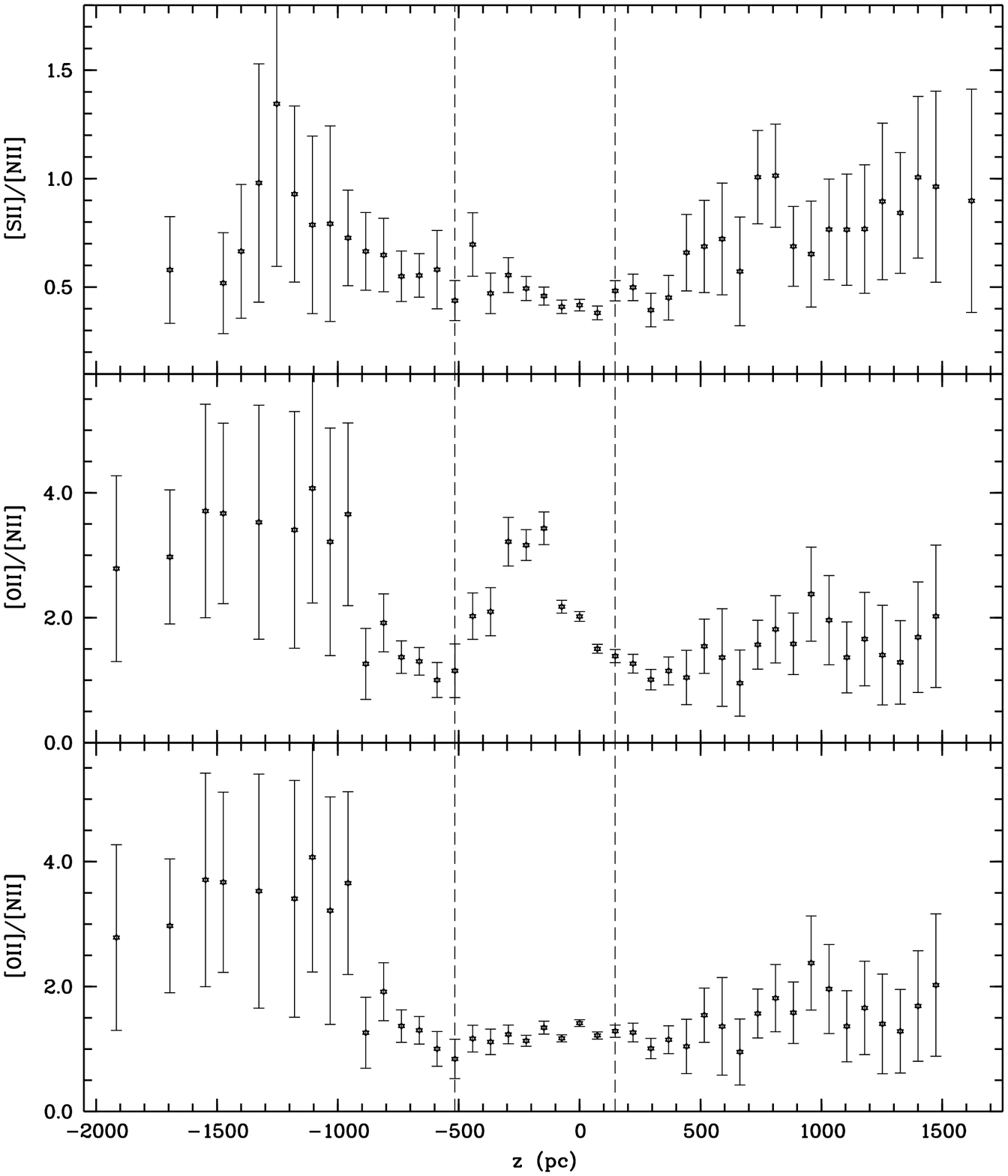}
\caption{\label{2f15}
Same as Fig. \ref{2f12}, but for NGC\,4634 (disk slit). Note that the data have
been smoothed by a running average over 9 rows of spectrum (590\,pc, see Table
\ref{2t2}) at $|z|>700$\,pc, but we plot the results for each row to show the
full extent of these data.}
\end{figure}

\begin{figure}
\epsscale{1.0}
\plotone{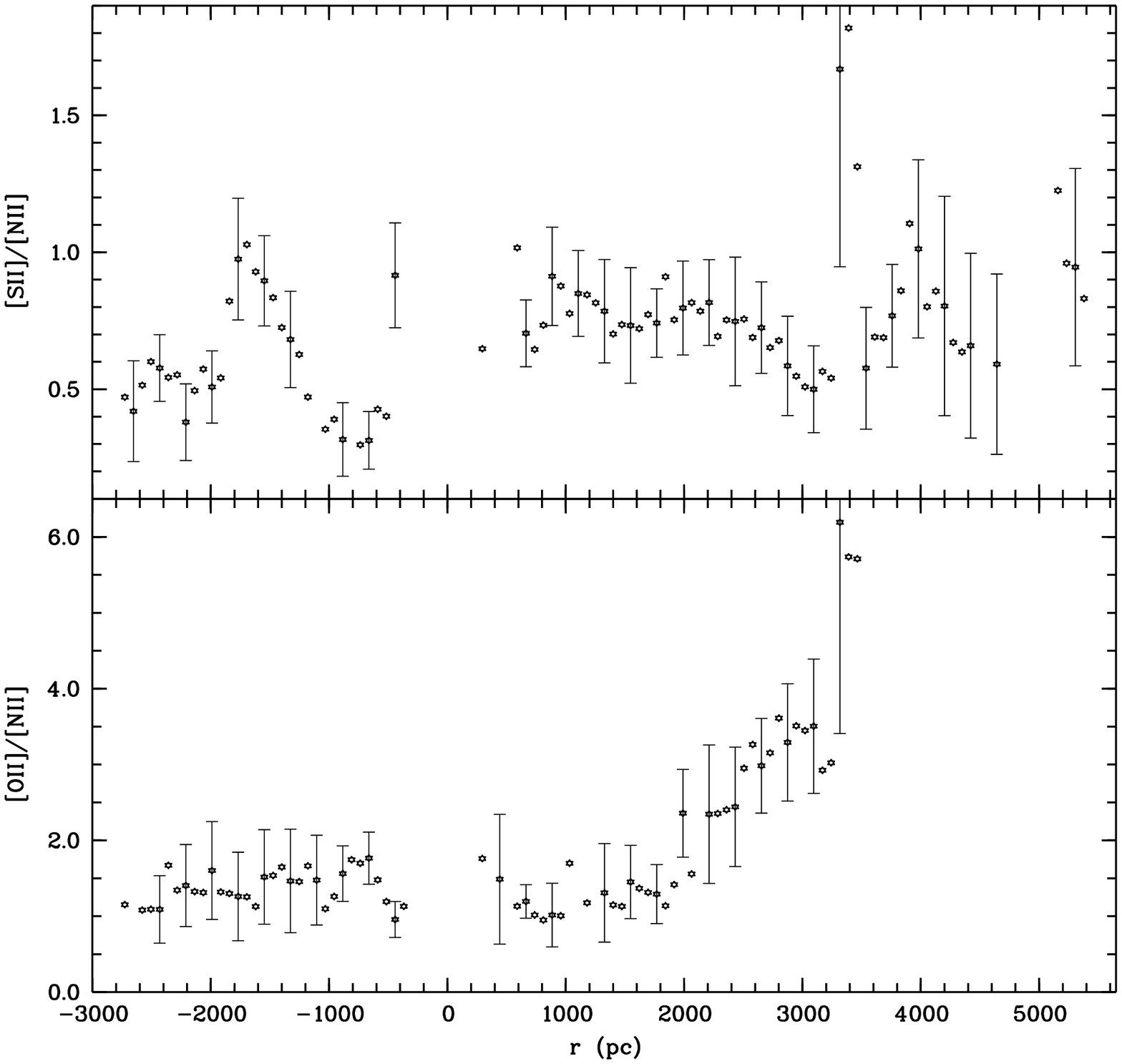}
\caption{\label{2f16}
Same as Fig. \ref{2f13}, but for NGC\,4634 (DIG slit). Note that the data have
been smoothed by a running average over 9 rows of spectrum (590\,pc, see Table
\ref{2t2}), but we plot the results for each row to show the full extent of
these data. Error bars are plotted only for every 3rd row of the spectrum.}
\end{figure}

\begin{figure}
\plotone{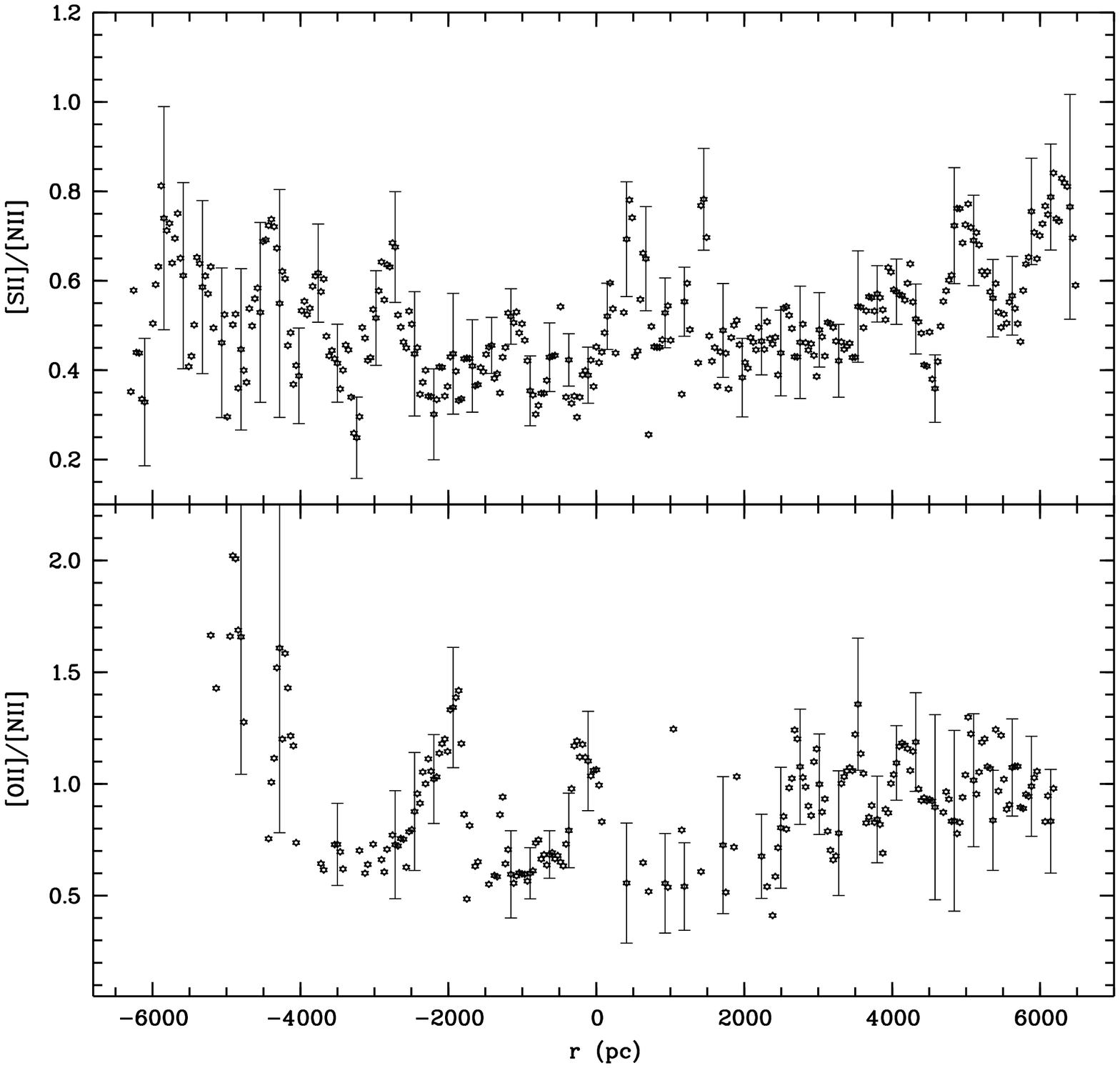}
\caption{\label{2f17}
Same as Fig. \ref{2f13}, but for NGC\,891. Note that the data have been smoothed
by a running average over 9 rows of spectrum (300\,pc, see Table \ref{2t2}), but
we plot the results for each row to show the full extent of these data. Error
bars are plotted only for every 7th row of the spectrum.}
\end{figure}

Pure photoionization models predict an increase in [\ion{S}{2}]/[\ion{N}{2}]
with increasing $|z|$. For example, an increase of $>0.2$ in
[\ion{S}{2}]/[\ion{N}{2}] in matter bounded models is correlated with an
increase in [\ion{S}{2}]/H$\alpha$ from 0.2 to 0.5 \citep{2rand} or even just
from 0.2 to 0.3 \citep{2dom}. Most variations in [\ion{S}{2}]/[\ion{N}{2}]
observed in our galaxies (including Paper I) appear to have lower gradients than
those predicted by photoionization models. In Figures \ref{2f12}--\ref{2f17}, we
show the [\ion{S}{2}]/[\ion{N}{2}] and [\ion{O}{2}]/[\ion{N}{2}] line ratios for
our five galaxies. The only disk slits in this paper are for
\objectname[]{NGC\,4634} and \objectname[]{NGC\,5775}. In
\objectname[]{NGC\,4634}, [\ion{S}{2}]/[\ion{N}{2}] increases with increasing
distance from the midplane. While the gradient appears to become smaller at
$z>+500$\,pc, the line ratio increases at first on the other side of the disk
and then seems to decrease with $|z|$ at $z<-1.3$\,kpc. However, given the
uncertainty in the data points, the [\ion{S}{2}]/[\ion{N}{2}] line ratio can
also be constant on this side. Unfortunately, our analysis does not extend that
far out into the halo due to the weak H$\beta$ line emission. It is not clear
what causes these variations (if they are real) or whether these changes are
small scale variations similar to those observed in the other slit positions.
From the disk slits of \objectname[]{NGC\,3079}, \objectname[]{NGC\,4631}, and
\objectname[]{NGC\,891} (Paper I) we know that [\ion{S}{2}]/[\ion{N}{2}] likely
has a minimum near the midplane, but becomes almost constant with small scale
variations at higher $|z|$. In \objectname[]{NGC\,5775},
[\ion{S}{2}]/[\ion{N}{2}] increases with increasing distance from the disk up to
about $z=-3$\,kpc, where it starts to decrease again. Both groups TDSUR and CR
observed strong variations in [\ion{S}{2}]/[\ion{N}{2}] with widths of up to
2\,kpc and heights of up to 0.5. It is therefore impossible to judge whether the
decrease beyond $z=-3$\,kpc is part of a local minimum, although the overall
trend in TDSUR's measurements is an increase with $|z|$ near the midplane and a
slight overall decrease with $|z|$ beyond $|z|=2$\,kpc. Our analysis extends up
to $z=-2$\,kpc. Due to the uncertainty of the extinction correction, it is not
clear whether a decrease in nitrogen abundance or an increase in the sulfur
ionization fraction causes the initial increase of [\ion{S}{2}]/[\ion{N}{2}] on
this side of the disk. The data points on the other side ($z>0$) are too sparse
to reveal any trends.

The slits parallel to the galactic disks in our sample show a relatively clear
trend in the [\ion{S}{2}]/[\ion{N}{2}] line ratio. In three of the four DIG
slits (\objectname[]{NGC\,3079}, \objectname[]{NGC\,4631},
\objectname[]{NGC\,891}), [\ion{S}{2}]/[\ion{N}{2}] increases with increasing
radial distance from the galactic center (although the line ratio decreases
again in \objectname[]{NGC\,4631} at $|r|>3-4$\,kpc, and the increase in
\objectname[]{NGC\,891} at $r<0$ is less obvious due to the larger
uncertainties). According to our analysis in section \ref{2s3}, this increase
appears to be dominated by a radial decrease in nitrogen abundance in these
galaxies. However, it is possible that variations in N$^+$/N also play a role
and thus weaken the influence of the nitrogen abundance as mentioned earlier.
Only in \objectname[]{NGC\,4634}, [\ion{S}{2}]/[\ion{N}{2}] stays more or less
constant with several smaller scale variations along the disk. Unfortunately, we
were not able to apply our analysis method to this slit position.

The behavior of the observed [\ion{O}{2}]/[\ion{N}{2}] line ratios varies
between slit positions. In the two disk slits, [\ion{O}{2}]/[\ion{N}{2}] seems
to increase with $|z|$ at first and then either levels off or undergoes small
scale variations. In the DIG slits of \objectname[]{NGC\,3079} and
\objectname[]{NGC\,4631}, [\ion{O}{2}]/[\ion{N}{2}] follows the increases in the
[\ion{S}{2}]/[\ion{N}{2}] line ratio confirming that the decreasing nitrogen
abundance is the cause for the observed increases in the DIG slits. This
correlation might also exist in \objectname[]{NGC\,891}, but is not obvious due
to the larger uncertainties. However, the [\ion{O}{2}]/[\ion{N}{2}] line ratio
parallel to the disk of \objectname[]{NGC\,4634} stays constant on one side
($r<0$), but increases with $r$ on the other side ($r>0$), where
[\ion{S}{2}]/[\ion{N}{2}] stays constant. Our data are insufficient to explain
this behavior. According to our analysis, the general behavior of
[\ion{O}{2}]/[\ion{N}{2}] can be explained by a combination of electron
temperature and nitrogen abundance. Little to no correlation can be found
between variations in [\ion{O}{2}]/[\ion{N}{2}] and changes in the oxygen
ionization fraction. The behavior of the line ratios [\ion{S}{2}]/[\ion{N}{2}]
and [\ion{O}{2}]/[\ion{N}{2}] in our small sample of galaxies (including the
slit positions in Paper I) rule out pure photoionization as source for the
observed line ratio variations.

In all our galaxies, the derived electron temperature increases with $|z|$ in
the disk slits perpendicular to the midplane and stays more or less constant
with increasing $|r|$ on a large scale in the DIG slits parallel to the midplane
(exception: increase in $T_4$ at $r>3$\,kpc in \objectname[]{NGC\,4631}). The
small scale variations in temperature derived from the model fits of the DIG
slits tend to be smaller than the vertical trends seen in the disk slits. Hence,
these vertical trends are influenced, but not dominated by possible small scale
structures. We observed the same behavior, i.e. increasing $T_4$ with increasing
$|z|$, in the disk slits of \objectname[]{NGC\,891}, \objectname[]{NGC\,4631},
and \objectname[]{NGC\,3079} (Paper I). Pure photoionization models always
require a decreasing ionization parameter with increasing height from the
midplane to obtain increasing line ratios of singly ionized species
\citep{2dom}. The decreasing ionization parameter also causes a decrease in
O$^{++}$/O and an increase in S$^+$/S with increasing $|z|$ \citep{2semb}. An
increasing electron temperature towards the halo appears to require an
additional heating mechanism. If a harder source of ionization caused the
observed increasing temperatures towards the halo, O$^{++}$/O would have to
increase as well. This does not seem to be so in \objectname[]{NGC\,3079} or
\objectname[]{NGC\,4634}, but may be the case in \objectname[]{NGC\,4631} or
\objectname[]{NGC\,891} (Paper I). However, with a harder source of ionization,
S$^+$/S decreases, as more sulfur becomes doubly ionized. Our data, including
Paper I, do not yield a coherent picture regarding both ionization fractions
(O$^{++}$/O and S$^+$/S) and the electron temperature, if a harder spectrum is
assumed to be the cause for the increase in electron temperature.

\begin{figure}
\plotone{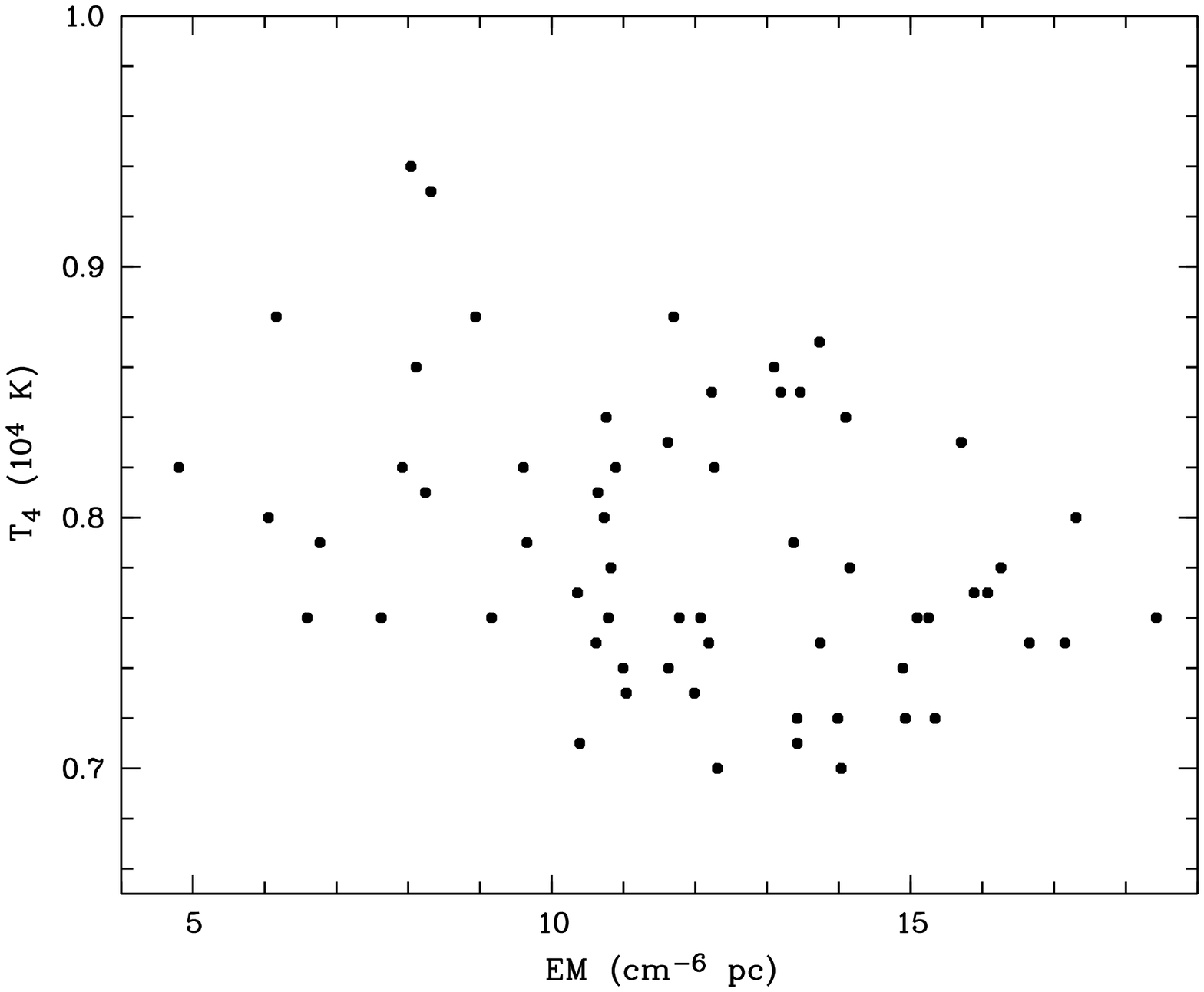}
\caption{\label{2f18}
Maximum allowed temperature $T_4$ versus emission measure EM for NGC\,3079 (DIG
slit). The electron temperature increases with decreasing emission measure,
suggesting that increases in temperature are associated with decreases in the
gas density.}
\end{figure}

It has been suggested that increases in temperature may be associated not only
with increases in $|z|$, but more generally with decreases in density $n$
\citep{2reyn}. Slits perpendicular to the disk might probe processes associated
with height from the midplane causing an increase in temperature with $|z|$ that
is not related to decreases in density. However, slits parallel to the midplane
eliminate this ambiguity and thus allow us to analyze whether changes in
temperature and variations in density are directly related. The intensity of the
H$\alpha$ emission line is proportional to $T_4^{-0.9}$ and the emission measure
EM, which depends on the product of $n^2$ and the pathlength $s$. Figure
\ref{2f18} shows a $T_4$ versus ${\rm EM}\propto I_{\rm H\alpha} T_4^{0.9}$ plot
for \objectname[]{NGC\,3079} (DIG slit). Since temperatures close to the maximum
allowed temperature yielded the best predictions according to our analysis in
section \ref{2s3}, only the maximum allowed temperature is used in the plot.
Despite the large scatter of the data points, a correlation between electron
temperature and the emission measure can be seen. The temperature increases with
decreasing emission measure. One does not expect the pathlength to change
significantly along most part of a parallel slit. It is also difficult to
imagine a mechanism that would relate the pathlength of the emitting gas to the
temperature. Thus, it is reasonable to assume that changes in EM are caused by
changes in density, more precisely by variations in $n^2$, and hence increasing
temperature is associated with decreasing density.

We found the same correlation in all our slits (DIG as well as disk slits). The
minimum allowed temperature stays more or less constant with decreasing
intensity in the DIG slits and increases in the disk slits. However, the minimum
allowed temperature did not produce good emission line predictions in our
analysis and therefore does not seem to represent the real temperature as well
as the maximum allowed temperature. The fact that in general the line ratio
predictions for the maximum allowed temperature match the observed line ratios
better than the predictions using minimum allowed temperature can be explained
at least partially by the extinction correction. Since we overcorrected for
extinction, the [\ion{O}{2}] emission became too strong which then required
higher temperatures. However, it is not clear whether other circumstances factor
into this behavior as well.

More correlations and anti--correlations might exist among the derived
quantities containing additional information on the processes in the DIG. While
most of these (anti--)correlations would be the result of (anti--)correlations
among the observed emission line ratios, some may be due to problems with the
model or the data. We have not yet examined the possible (anti--)correlations
among the other derived quantities, as our goal in this paper was to analyze
trends in the electron temperature with regard to the emission line ratios.

\section{CONCLUSIONS}

We observed five nearly edge--on galaxies to examine variations in their
emission line ratios. The forbidden line intensities relative to H$\alpha$ are
higher in regions of lower line emission. Since this behavior has been observed
in other galaxies as well covering a range of types and properties, this trend
seems to be associated with the gas itself rather than with peculiar situations
within a given galaxy. In most cases the observed variations in the line ratios
are consistent with systematic changes in electron temperature within the
ionized gas and with normal radial chemical abundance gradients. In particular,
the electron temperature seems to increase with increasing distance from the
midplane of the galaxies. The derived values for electron temperature,
ionization fractions, and nitrogen abundance rule out pure photoionization as
the explanation for the observed line ratios and suggest that the increase in
electron temperature with increasing $|z|$ is not necessarily caused by an
additional source of ionization, but by an additional heating mechanism which
does not ionize the diffuse gas (see Haffner et al. 1999 and references
therein). Moreover, our derived temperature variations along the slits seem to
be correlated with variations in density, with regions of lower density having
higher temperatures.

A definitive measurement of the electron temperature in the eDIG would provide a
critical test of our model. Detection of increasing electron temperatures at low
densities would establish that eDIG emission partially results from the
operation of heating sources other than photoionization in the gaseous halos of
galaxies. Given the low intensities of eDIG optical emission lines, such direct
measurements are extremely difficult, but may be feasible through emission line
velocity width measurements in the Milky Way (e.g. Reynolds 1985).

We also conclude that it is important to gain a better understanding of the
extinction in galaxies, as the applied extinction correction seems to
overcorrect the blue emission lines and can significantly change the values of
derived abundances, ionization fractions and the electron temperature. One
possibility would be to obtain spectra of sufficient signal--to--noise to be
able to measure the Balmer lines H$\gamma$ and H$\delta$ (which were visible in
two or three of our galaxies) and to derive the galactic extinction from these
Balmer decrements.

\acknowledgements

B. O. is thankful to the staff on Kitt Peak and to J. Hoffman for their support
during the observations. She is also grateful to S. Jansen for his computer
support. The authors thank the anonymous referee for his comments and suggestions resulting in an improved paper. This research was funded by the NSF through
grant AST96--19424 and in part by the Graduate School of the University of
Wisconsin--Madison. The observing run at the 2.1\,m was funded in part by the
National Optical Astronomy Observatory. The Digitized Sky Surveys were produced
at the Space Telescope Science Institute under U. S. Government grant NAGW-2166.
The images of these surveys are based on photographic data obtained using the
Oschin Schmidt Telescope on Palomar Mountain. The National Geographic Society --
Palomar Observatory Sky Atlas was made by the California Institute of Technology
with grants from the National Geographic Society. The Second Palomar Observatory
Sky Survey was made by the California Institute of Technology with funds from
the National Science Foundation, the National Geographic Society, the Sloan
Foundation, the Samuel Oschin Foundation, and the Eastman Kodak Corporation. The
Oschin Schmidt Telescope is operated by the California Institute of Technology
and Palomar Observatory. Supplemental funding for sky--survey work at the STScI
is provided by the European Southern Observatory.

\clearpage

\begin{figure}
\figurenum{6b}
\plotone{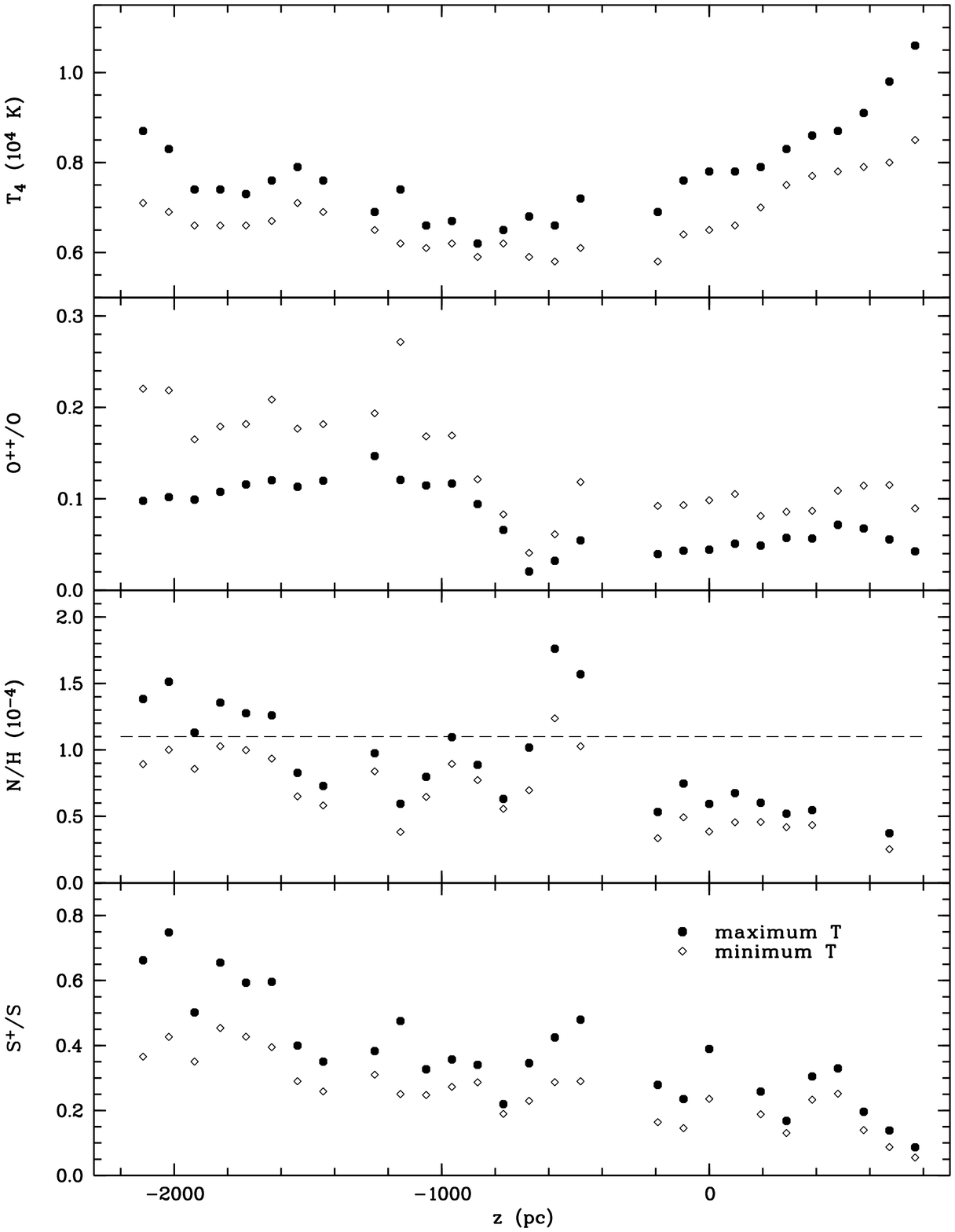}
\caption{}
\end{figure}
\begin{figure}
\figurenum{6c}
\plotone{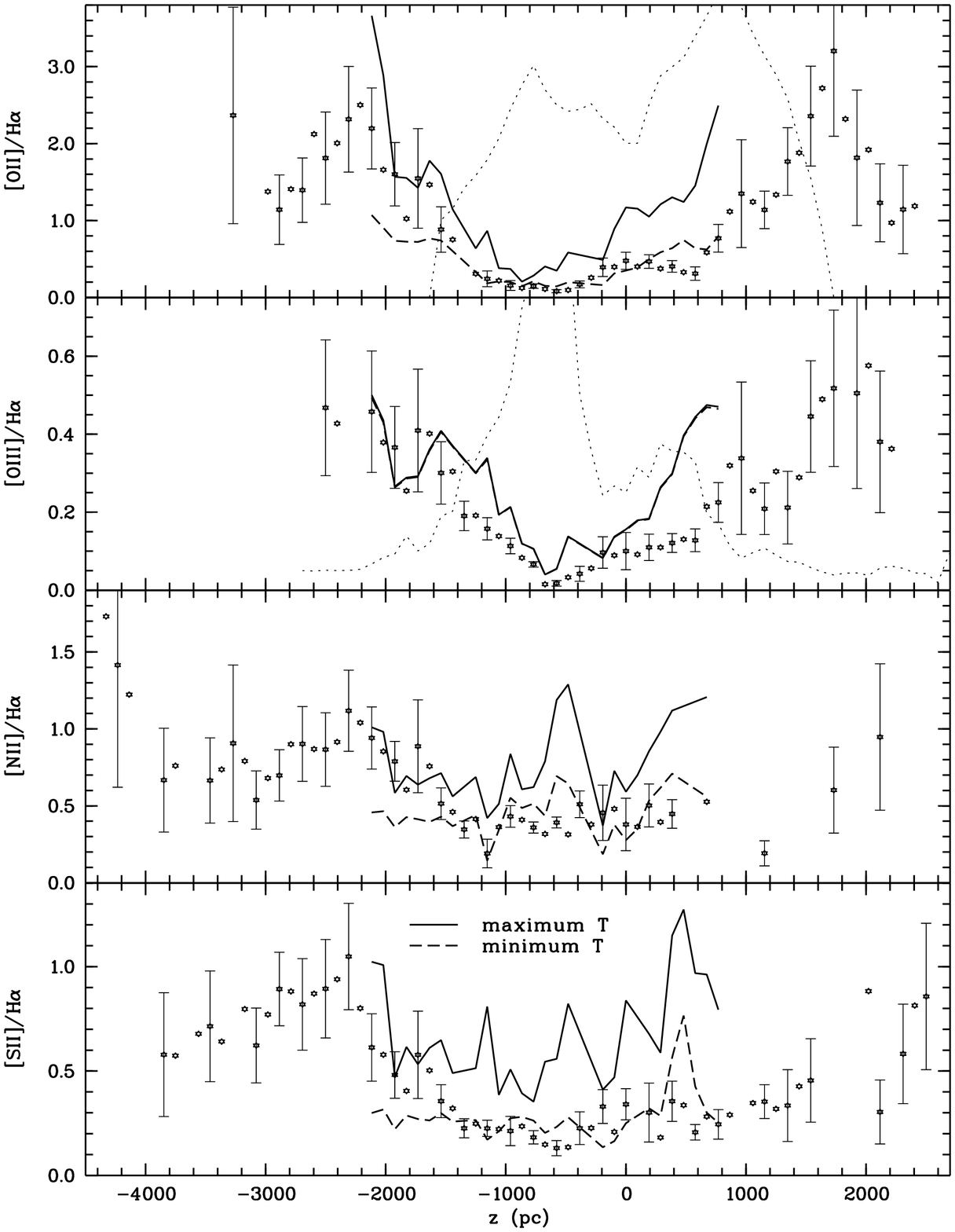}
\caption{}
\end{figure}
\begin{figure}
\figurenum{6d}
\plotone{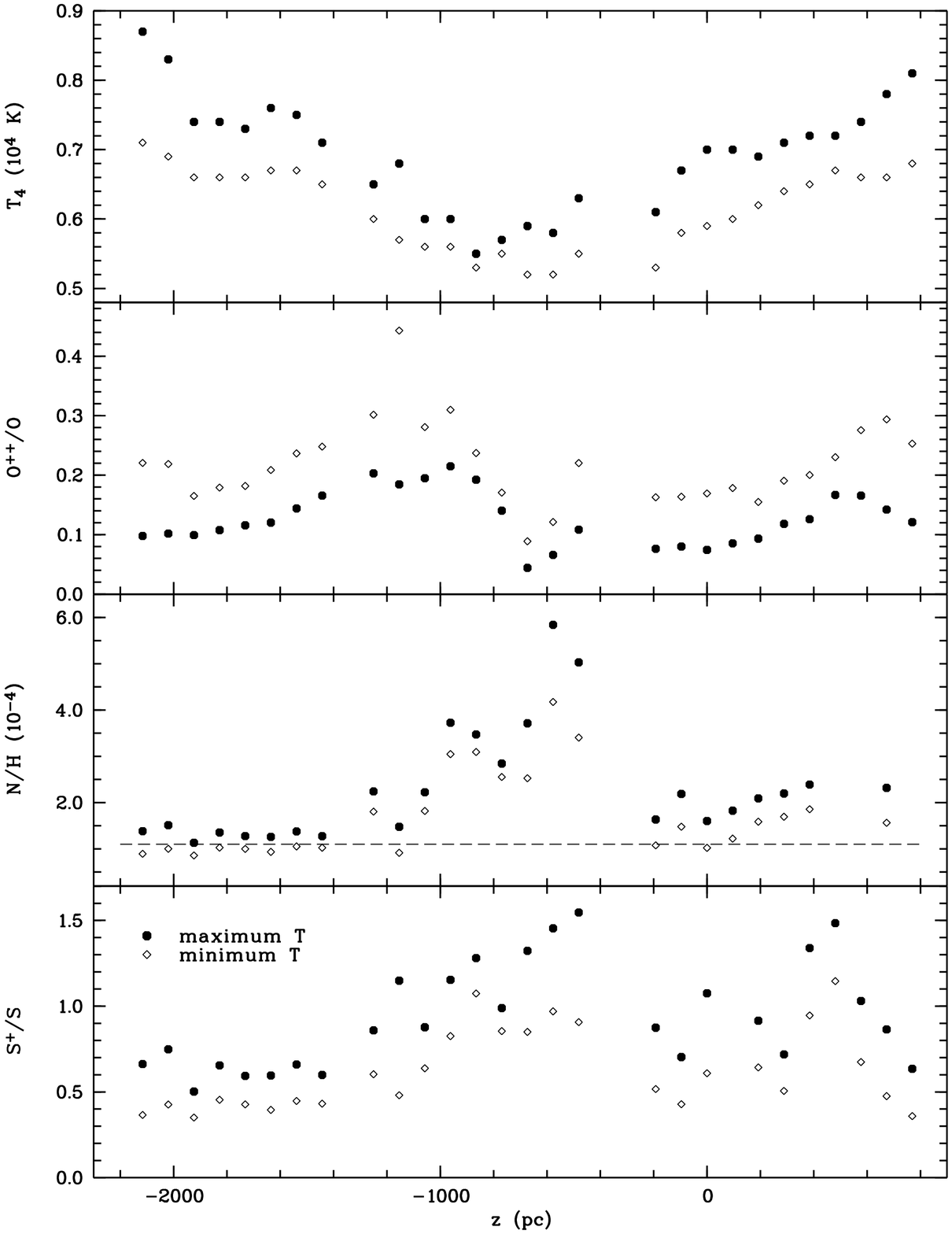}
\caption{}
\end{figure}
\begin{figure}
\figurenum{7b}
\plotone{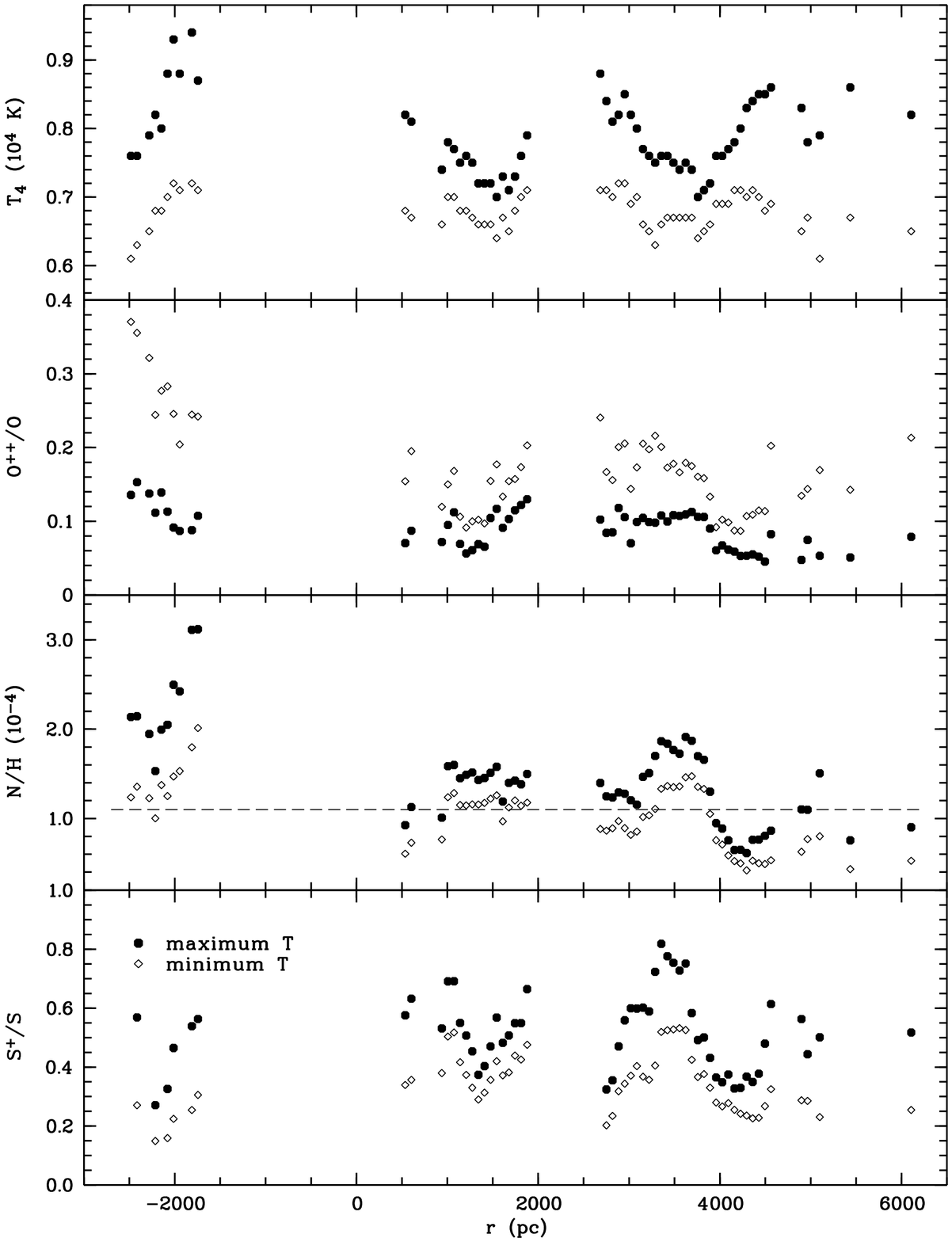}
\caption{}
\end{figure}
\begin{figure}
\figurenum{8b}
\plotone{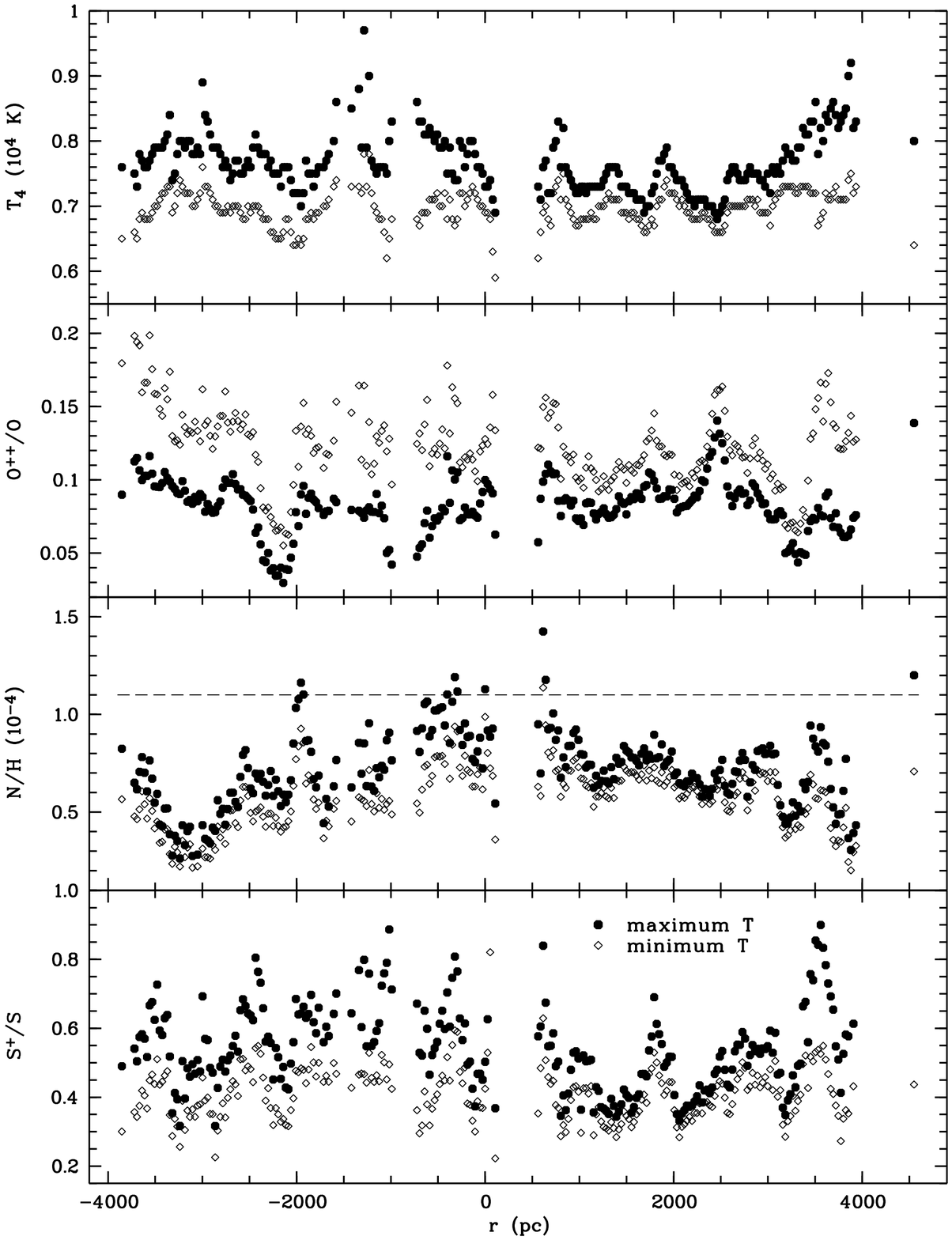}
\caption{}
\end{figure}
\begin{figure}
\figurenum{9b}
\plotone{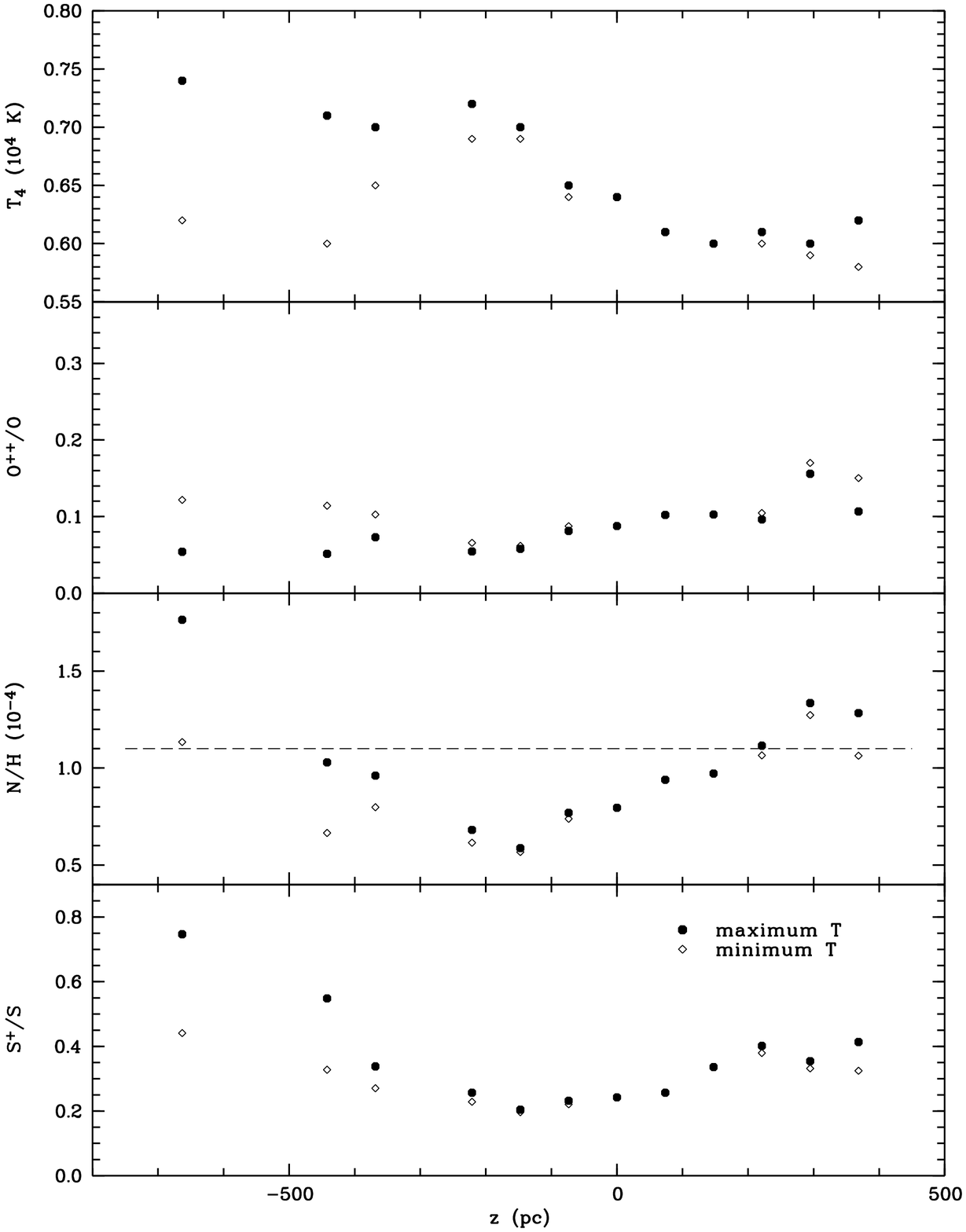}
\caption{}
\end{figure}
\begin{figure}
\figurenum{9c}
\plotone{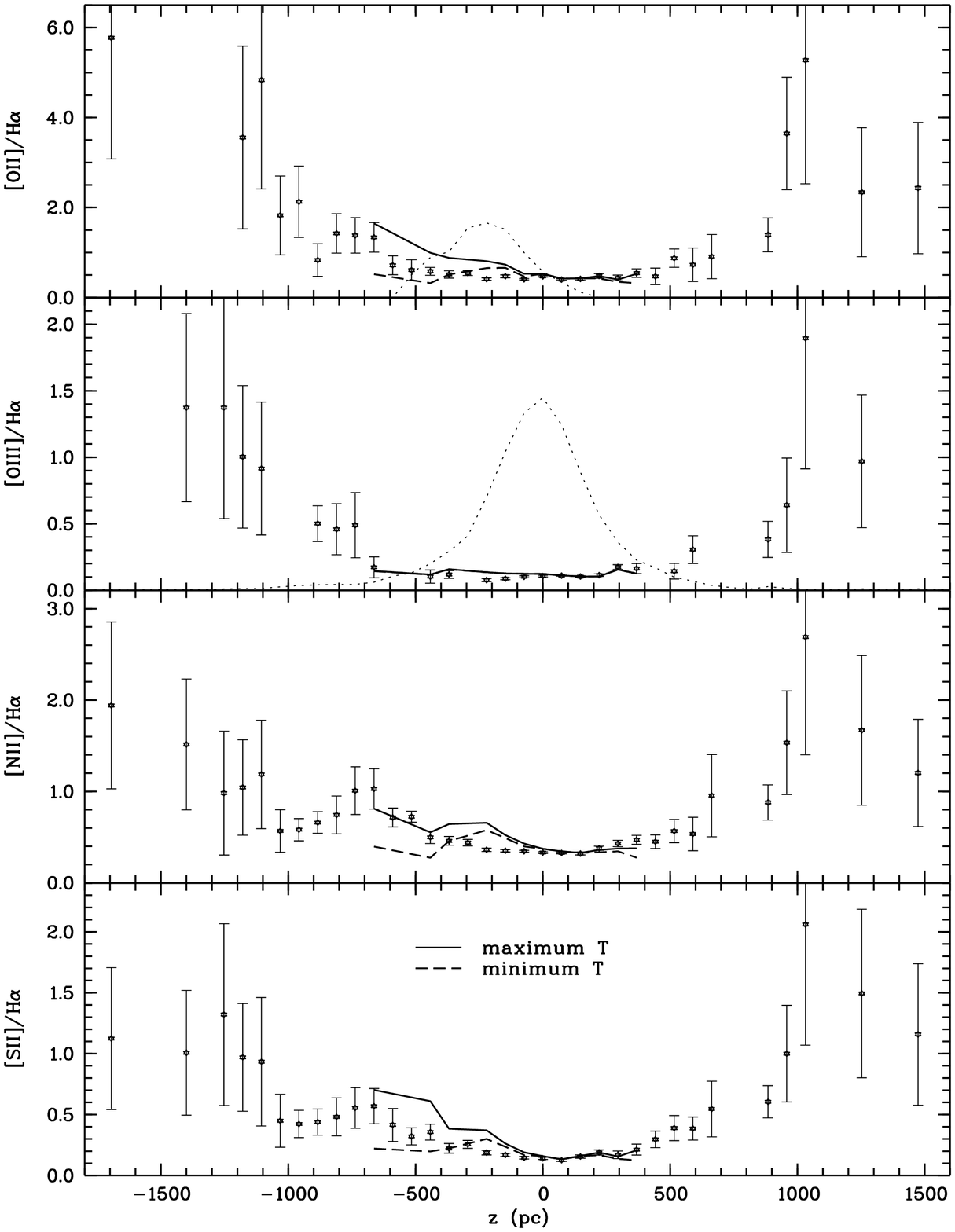}
\caption{}
\end{figure}
\begin{figure}
\figurenum{9d}
\plotone{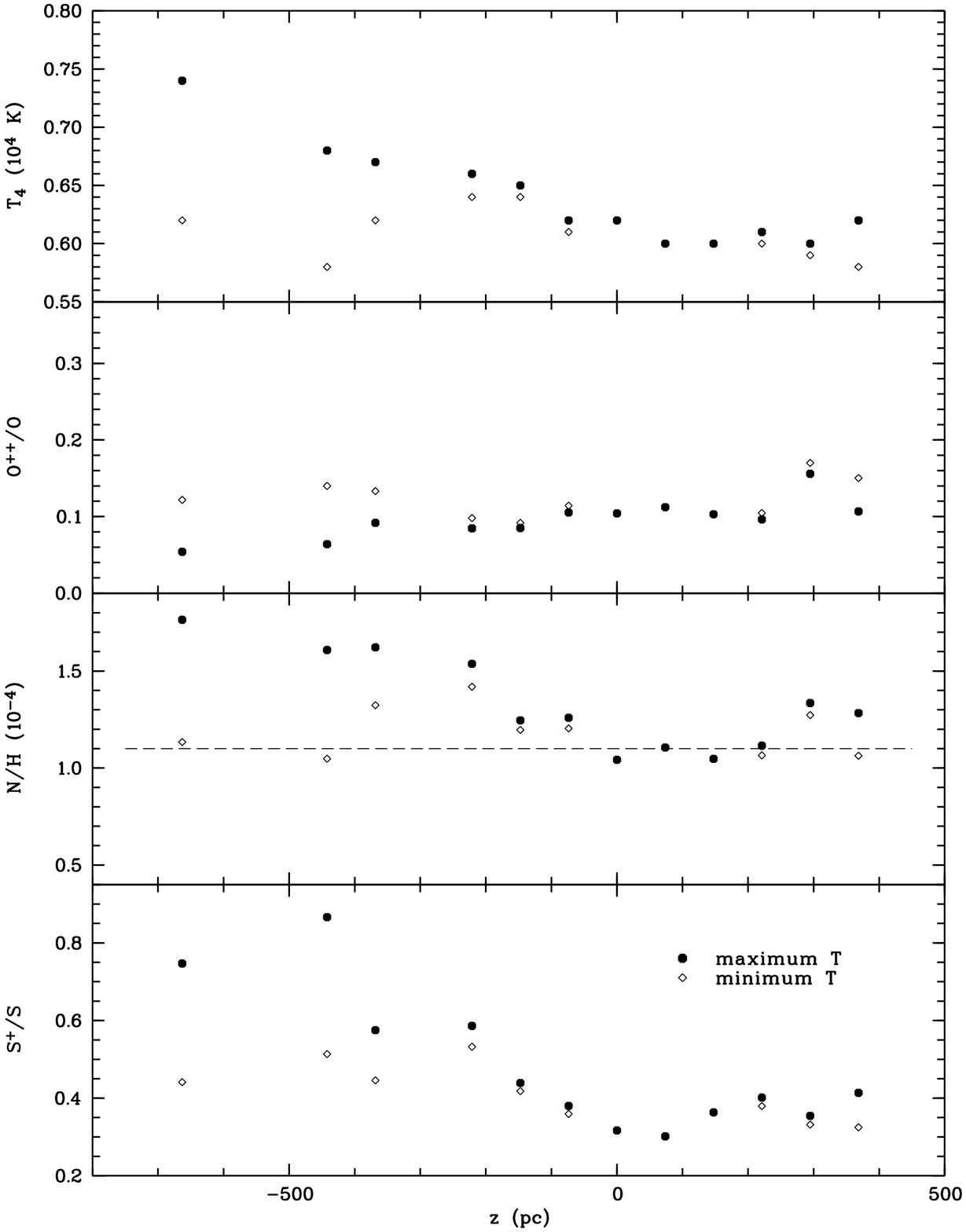}
\caption{}
\end{figure}
\clearpage

\end{document}